\documentclass{aa}
\usepackage{psfig,times}
\def\X21{PKS~2149$-$306\/}
\def\P2126{PKS~2126$-$158\/}
\def\1442{Q~1442+2931\/}
\def\Q0{Q~0000$-$263\/}
\def\etal{et al.\/ }
\def\ergs{erg\,cm$^{-2}$\,s$^{-1}$}

\def\la{~\raise.5ex\hbox{$<$}\kern-.8em\lower 1mm\hbox{$\sim$}~}
\def\ma{~\raise.5ex\hbox{$>$}\kern-.8em\lower 1mm\hbox{$\sim$}~}
\begin{document}
\title{XMM-Newton observations of four high-z quasars}
\author{ E. Ferrero\inst{1} and W. Brinkmann\inst{2}}
\offprints{E. Ferrero; e-mail: ferrero@mpe.mpg.de}
\institute{Max--Planck--Institut f\"ur extraterrestrische Physik,
  Postfach 1312, D-85741 Garching, FRG
\and Centre for Interdisciplinary Plasma Science,
 Max--Planck--Institut f\"ur extraterrestrische Physik,
 Postfach 1312, D-85741 Garching, FRG}
\date{Received ?; accepted ?}
\abstract{
We present the results of XMM-Newton observations of  four high-z quasars, two radio-loud
and two radio-quiet. One of the radio-loud objects, PKS 2126$-$158, clearly shows absorption 
in excess of the galactic value  as claimed before from ASCA and ROSAT observations.
For  PKS 2149$-$306  the evidence for excess 
absorption is only marginal  in contrast to previous results. 
The location of the absorber in  PKS 2126$-$158 is compatible with the 
redshift of the source.  
Both, a warm and a cold absorber are allowed by the X-ray data.
Both {quasars} have very flat photon spectra ($\Gamma \la 1.5$) and
the high photon statistics reveal small deviations from a simple power law form.
For the two radio-quiet objects, Q 0000$-$263 and Q 1442+2931, we determine, for
the first time, reliable spectral parameters. Both quasars have steeper power laws
($\Gamma \sim 2$) and show absorption consistent with the galactic value, similar to 
radio-quiet quasars at low redshifts.
   In the case of  Q 0000$-$263 the presence of the damped Ly$\alpha $ system 
($N({\rm HI})\sim 2.6\times 10^{21}$ cm$^{-2}$ at z = 3.39) cannot be examined further  due to the 
limited photon statistics in all instruments.
}

\titlerunning{XMM observations of high-z quasars}
\authorrunning{E. Ferrero \& W. Brinkmann}
\maketitle

\section{Introduction}

High redshift quasars are the most luminous and most distant,
continuously emitting sources of electromagnetic radiation 
in the observable Universe.
In particular at X-ray energies luminosities greater than 10$^{\rm 47}$ erg
s$^{\rm -1}$ are observed for some objects (Brinkmann \etal 1995).
These extreme luminosities imply the presence
of super-massive black holes (M$_{\rm BH} \sim 10^9 $M$_\odot$) in the
centers  of the sources and therefore provide severe constraints on 
theories of structure formation in the early Universe. 
High redshift quasars are key objects to understand the cosmological
evolution of the physical properties in and around quasars 
on the longest possible time scales and they present rare,
directly observable tracers of the physical conditions of the
early Universe.
  
The X-ray band is important for at least two reasons: the X-ray luminosity
represents a large fraction of the bolometric luminosity of quasars
(Elvis \etal 1994a)  and it is known from variability and spectral
studies (e.g. Mushotzky et al. 1993, Yuan et al. 1998a,
Yuan \& Brinkmann 1998) that the X-rays
are emitted very close to the central engine - a region which is not accessible
at any other wavelength with current instrumentation.
  
To understand quasars a detailed knowledge is required about
the mechanisms of the quasar emission and
about the cosmical evolution of the objects.
Radio, optical, and X-ray luminosity functions indicate that the typical 
luminosity of a quasar at $z$ $\sim$ 3  is higher by a factor of
50$-$100 compared to a local one (Boyle \etal 1993), but the result
is dominated by a few high redshift quasars with extreme luminosities.
 How are these enormous amounts of radiative energy produced and do
these properties evolve with time? 
What is the origin of the radio-loud / radio-quiet dichotomy, i.e., the fact
that about 10\% of the quasars show significant radio emission, are
brighter at high energies and more variable?

 Early studies of low-z quasars by the {\it Einstein}
observatory (Zamorani et al. 1981, Wilkes \& Elvis 1987) 
and ROSAT (Brinkmann et al. 1997, Yuan et al. 1998a) revealed
 X-ray  luminosities  up to
$\sim 10^{47}$ erg s$^{-1}$, which are roughly proportional to the 
corresponding optical luminosities, with a large dispersion. 
The  X-ray spectra in the soft band can be  described 
by power laws with a wide range of slopes with averages 
 around $\Gamma \ma 1.5$ for radio-loud quasars and
$\Gamma \ma 2.0$ for radio-quiet quasars, both flattening with
increasing redshift. 
In the harder  ASCA energy band 
radio-loud quasars have approximately $\Gamma \sim 1.6$ and
radio-quiet quasars  $\Gamma \sim 1.9$  (Reeves \& Turner 2000).
At a given optical luminosity radio-loud
quasars are typically $\sim 3$ times more X-ray luminous than radio-quiet
quasars.
A measure for this luminosity ratio is the  X-ray loudness $\alpha _{\rm ox}$,
the broad spectral index of a nominal power law from the optical to
the X-ray band.
From {\it Einstein} and ROSAT observations (Zamorani et al. 1981, 
Wilkes \& Elvis 1987, Brinkmann et al. 1997, Yuan et al. 1998a) it is seen
that this quantity is smaller for radio-loud quasars 
($\alpha _{\rm ox, rl}\sim 1.25$) than for radio-quiet quasars
($\alpha _{\rm ox, rq}\sim 1.6$).
The observationally found  dependence of $\alpha _{\rm ox}$ 
on redshift and optical luminosity, where the primary correlation is
that on optical luminosity (Avni \& Tananbaum 1982, 1986, Wilkes et al. 1994, 
Yuan et al. 1998a), would imply a different evolution in the optical and
X-ray regime. However, Brinkmann et al. (1997) and Yuan et al. (1998b)
argue that this dependence is not a physical property of the population
but can be introduced by selection effects and the luminosity
dispersions of the samples in the optical and X-ray band.
  
The differences between the two classes of quasars
can be  explained in a two component  emission model
(Zamorani et al. 1981, Wilkes \& Elvis 1987), 
where a steep soft component, linked to the optical 
emission, is present in all quasars  and a second
flat spectrum  hard component linked to the radio emission through the SSC
mechanism dominates the  X-ray emission of radio-loud quasars.

The question of whether quasars do exhibit spectral evolution or not is 
fundamental and has direct impact on quasar formation models.
The question of whether the observed difference between radio-quiet and
radio-loud quasars at low redshifts  persists
to high redshifts cannot be answered conclusively with current data, which 
tentatively indicate that no evolution takes place in radio-loud quasars,
but the number of objects is still very low. Spectra for high redshift
radio-quiet quasars with reasonable quality hardly exist.
A study of the 
currently unknown spectral properties of high redshift radio-quiet quasars 
is also relevant for our understanding of the cosmic X-ray background and 
the contribution of these objects.

One of the major results of the ROSAT and ASCA observations of high redshift
radio-loud quasars was the detection of absorption in excess of what 
is expected from the galactic $N_{\rm H}$-value (Elvis \etal 1994b,
Siebert \etal 1996, Brinkmann \etal 1997, Cappi \etal 1997, Yuan \etal 2000).
 However, due to the low 
signal-to-noise of the spectra and the insufficient energy resolution of 
the instruments, it is impossible to unambiguously determine whether the 
absorption is galactic, inter galactic or intrinsic 
to the quasar and, in some cases, the absorption appears to be even temporarily
variable (Schartel \etal 1997). 
Current observations
tentatively indicate an intrinsic absorption site. Related to this is the
question, whether excess absorption is also a feature of radio-quiet 
quasars (Yuan \etal 1998a, Yuan \& Brinkmann 1998).
Up to now, the answer is no, but the data are far from being conclusive.
Any systematic differences in the spectral and/or absorption properties 
between radio-loud
and radio-quiet quasars can be extremely important for an 
understanding of the formation processes for quasars and the 
radio-loud/radio-quiet dichotomy.

Furthermore, if damped Ly$\alpha$ systems are in the line-of-sight, it should be
possible, with the help of X-ray observations, to determine the ionization 
state of these systems and thus derive limits on the size, temperature and
density for them (Fang \& Canizares 2000). However, so far 
current data do not give significant constraints (Fang \etal 2001).

In this paper we present the results of XMM-Newton observations of four
 high redshift quasars. Two are radio-loud, \P2126 ($z$ = 3.27) 
and \X21 ($z$ = 2.34) and two are radio-quiet, \1442 ($z$ = 2.64) and \Q0 ($z$ = 4.10).
All of them have been observed in X-rays before. From ASCA data
(as well as ROSAT data in {\bf the} case of \P2126) Cappi \etal (1997) claim 
excess absorption towards the radio-loud  objects.
No excess absorption towards the radio-quiet objects was  found 
from  ROSAT observations (Bechtold et al. 1994a, Reimers et al. 1995,
Kaspi et al. 2000).

\Q0 was observed by ROSAT in a PSPC pointed observation on November 30,
1991. From this observation Bechtold et al. (1994a) found an energy
index $\alpha =1.30\pm 0.23$ and no indications for extra absorption.  
From this observation and from a second PSPC pointing 
on November 26, 1991 Kaspi et al. (2000) determined  an unabsorbed 
flux in the $0.1-2.0$ keV band of $f_{\rm 0.1-2.0~keV}=6.5\times 10^{-14}$
\ergs and an optical - to - X-ray index  $\alpha _{\rm ox}=1.65$.
 However, they had to assume a photon index
$\Gamma =2.0$ and galactic absorption  as the data were insufficient
to perform a spectral fit.

The discovery of \1442 was reported by Sanduleak \& Pesch (1989) and the source
was first observed in X-rays by the ROSAT PSPC in November 1992 and
in July 1993 (Reimers et al. 1995). The accumulated net counts were not sufficient
to allow a spectral analysis,  however  Reimers et al. (1995) find no  
indications for excess absorption as the photons were distributed  
over the whole ROSAT energy band.
 
\P2126 was detected in X-rays by {\it Einstein} (Zamorani \etal 1981).
The ROSAT PSPC spectrum was presented in Elvis \etal (1994b) and
from ASCA observations Serlemitsos \etal (1994) constrained the 
redshift of the absorber at $z$ $<$ 0.4.
However, Cappi \etal (1997) could not reproduce these results and
attribute  them to the use of older response matrices  by 
Serlemitsos \etal (1994). 

With a 2$-$10 keV X-ray luminosity of $L_{\rm x} \sim 6\times 10^{47}$ erg~s$^{-1}$
\X21 is one of the most luminous radio-loud quasars in the Universe.
The ROSAT All Sky Survey data and   the ASCA observations of \X21
were first discussed in Siebert \etal (1996).
The absorption column density found by Cappi \etal (1997) is slightly higher,
but consistent, with that given by Siebert \etal (1996).
However, the extra absorption ( $\Delta N_{\rm H} \sim 4 \times 10^{20}$
cm$^{-2}$) is not large and the deduced value 
could be affected significantly by calibration
uncertainties of the SIS detectors.

The 2$-$10 keV flux of $8\times10^{-12}$ erg cm$^{-2}$ s$^{-1}$ reported
from a BeppoSAX observation in October 1997 (Elvis \etal 2000)  is
only $\sim$ 80\% of the  ASCA flux in 1994 (Cappi \etal 1997).
The hard power law index of $\Gamma = 1.4\pm0.05$  is similar to
the values found by ASCA ($\Gamma = 1.54\pm0.05$, Cappi \etal 1997),
however, the LECS showed an excess of counts
below 1 keV  and the absorption had to be fixed to the galactic
value. 
No evidence for a red-shifted Fe-K emission line was seen in the spectrum
and an upper limit of 63~eV for the equivalent width of a line at 
$\sim$ 5 keV claimed by Yaqoob \etal (1999) was given.

In recent {\it Chandra} observations Fang \etal (2001) do not find significant
excess absorption towards \X21. The source flux has decreased by about 30\%
compared to the ASCA observation and the deduced power law index  
$\Gamma = 1.255 \pm 0.020$ is significantly lower than the value of
$\Gamma \sim 1.55$ seen by ASCA. Further, the emission feature around 5~keV
reported by Yaqoob \etal (1999)  was not found in the {\it Chandra}  data.
 
In this paper we will present the  results of XMM-Newton observations
of these four quasars. 
In the next section we will give details on the observations  and discuss the 
temporal behavior of the objects.
We will then present the spectral analyses of the objects and discuss in section 4
their broad band properties, in particular the amount of absorption towards the
sources.
A general discussion and a summary will be given in section 5.

\section{The data}

\setcounter{table}{0}
\begin{table*}[t]
\small
\tabcolsep1ex
\caption{\label{data} { Data of observations}} 
\begin{tabular}{llcccc}
\noalign{\smallskip} \hline \noalign{\smallskip}
\multicolumn{1}{c}{Source} & \multicolumn{1}{c}{Observing date} &
\multicolumn{1}{c}{Instrument} &
\multicolumn{1}{c}{Mode} & \multicolumn{1}{c}{Filter } &
\multicolumn{1}{c}{Exposure} \\
\multicolumn{1}{c}{  } & \multicolumn{1}{c}{ } &
\multicolumn{1}{c}{  } &
\multicolumn{1}{c}{ } & \multicolumn{1}{c}{ } &
\multicolumn{1}{c}{(ksec)} \\
\noalign{\smallskip} \hline \noalign{\smallskip}
\X21 & May 1, 2001 & PN & Large Window & Medium & $\sim 21.8$ \\
 & & MOS1/2 & Full Window & Medium & $\sim 24$ \\ 
 & & RGS1/2 & Spectroscopy HER &  & $\sim 25$\\
 & & OM & Imaging & UVW2 & $\sim 20^{\dagger}$\\
\noalign{\smallskip} \hline \noalign{\smallskip}
\P2126 & May 1, 2001 & PN & Extended Full Window & Medium & $\sim 18$ \\
 & & MOS1/2 & Full Window & Medium & $\sim 22.8$ \\ 
 & & RGS1/2 & Spectroscopy HER &   & $\sim 23$\\
 & & OM & Imaging & Grism 1& $\sim 20^{\dagger}$\\
\noalign{\smallskip} \hline \noalign{\smallskip}
\Q0 & June 25, 2002 & PN & Extended Full Window & Thin & $\sim 43.5$ \\
 & & MOS1/2 & Full Window & Thin & $\sim 50.3$ \\
 & & RGS1/2 & Spectroscopy HER & & $\sim 23$ \\
\noalign{\smallskip} \hline \noalign{\smallskip}
\1442 & August 1, 2002 & PN & Extended Full Window & Thin & $\sim 37$ \\
 & & MOS1/2 & Full Window & Thin & $\sim 39 $ \\
 & & RGS1/2 & Spectroscopy HER &  & $\sim 42$ \\
 & & OM & Imaging & UVW2 & $\sim 20^{\dagger}$ \\ 
\noalign{\smallskip}\hline
\end{tabular}
\medskip

 $^{\dagger}$: Divided into five exposures of $\sim 4$ ksec each. \\
\end{table*}

 The observational details for the four sources are reported
in Table \ref{data}.
All PN and MOS data were reprocessed using XMMSAS version 5.3.0;  for the 
RGS data  XMMSAS version 5.3.3 has been used.

\begin{figure}
\psfig{figure=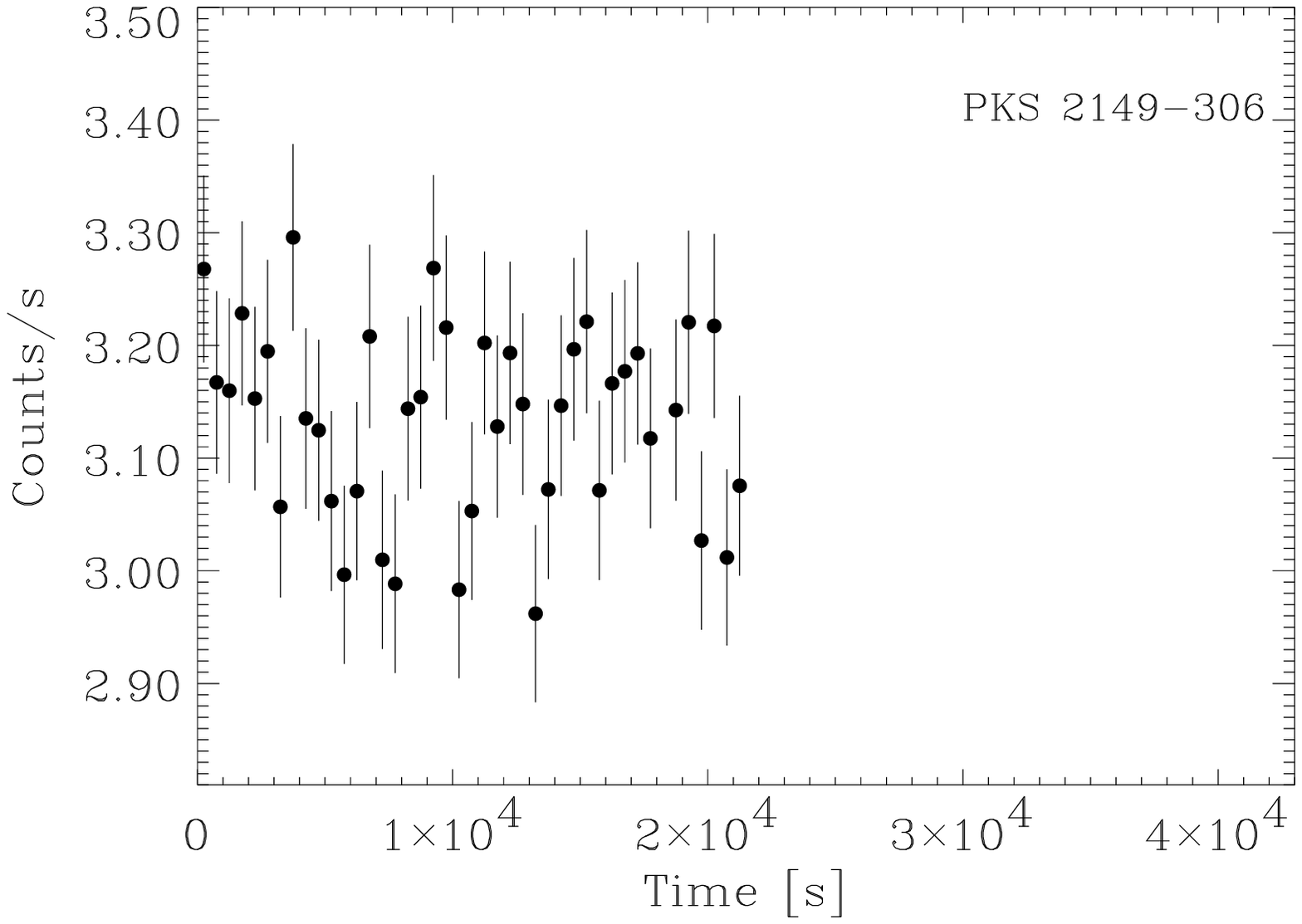,height=3.7truecm,width=8.5truecm,%
 bbllx=76pt,bblly=424pt,bburx=532pt,bbury=693pt,clip=}
\psfig{figure=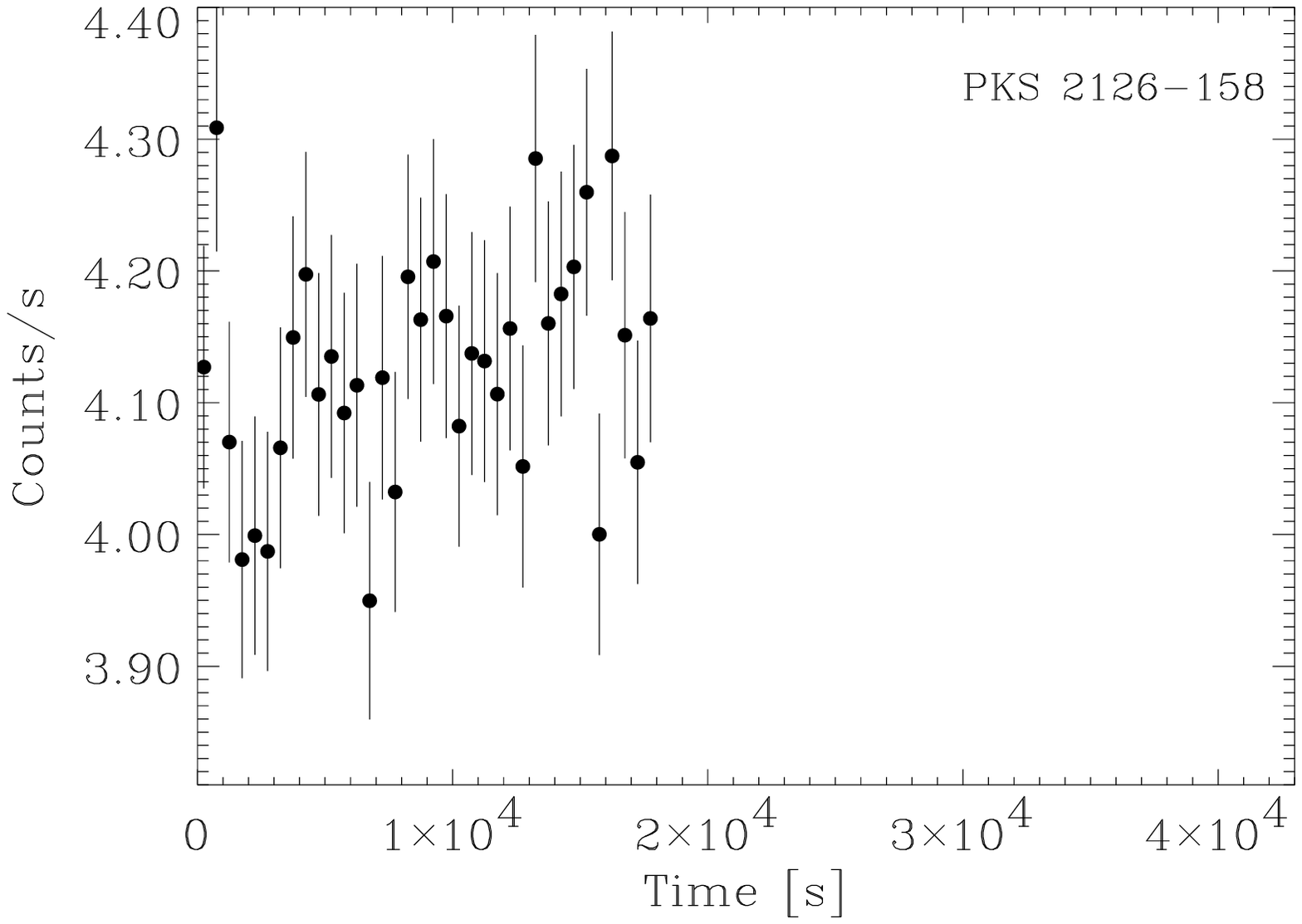,height=3.7truecm,width=8.5truecm,%
 bbllx=76pt,bblly=424pt,bburx=532pt,bbury=693pt,clip=}
\psfig{figure=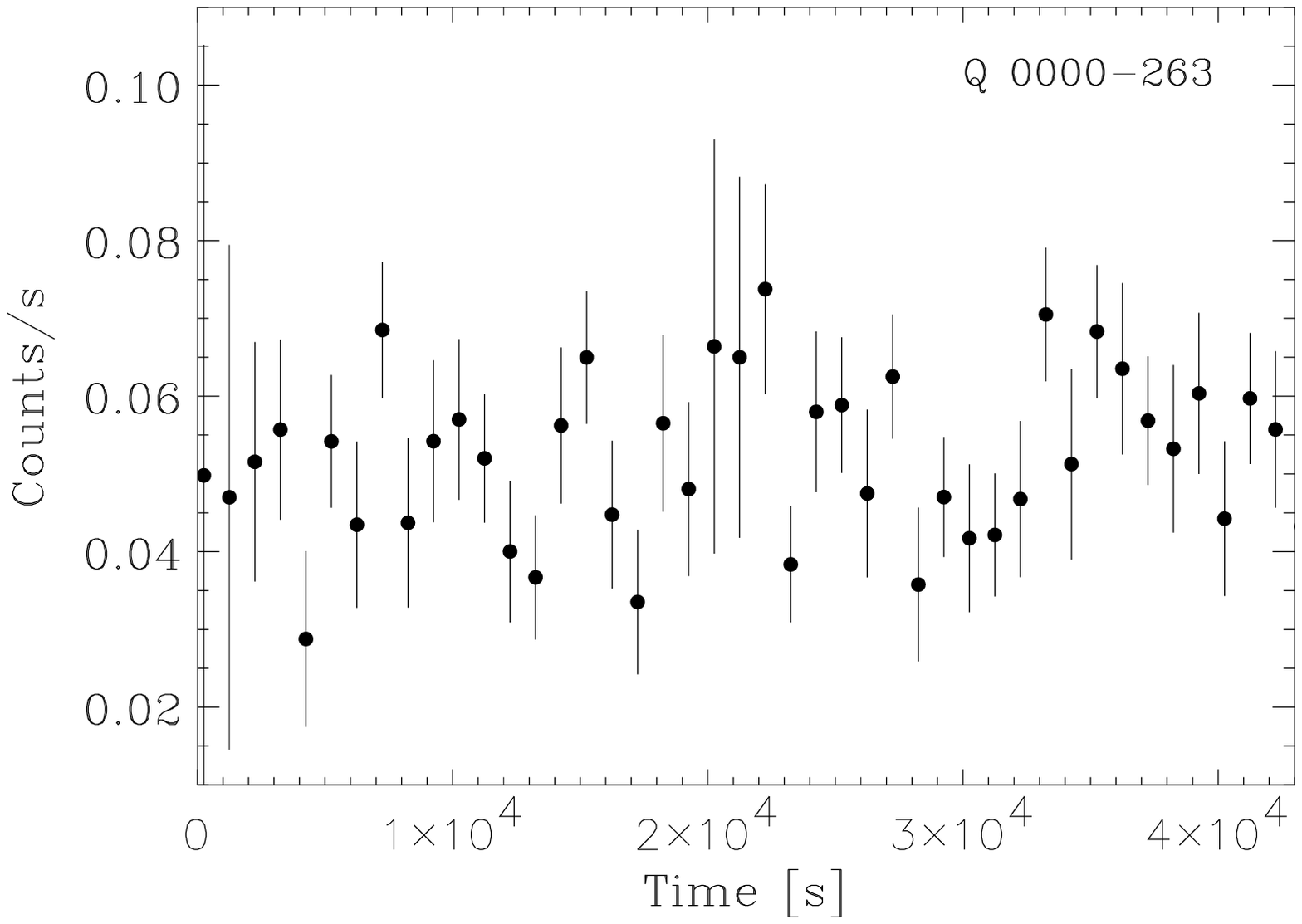,height=3.7truecm,width=8.5truecm,%
 bbllx=76pt,bblly=424pt,bburx=532pt,bbury=693pt,clip=}
\psfig{figure=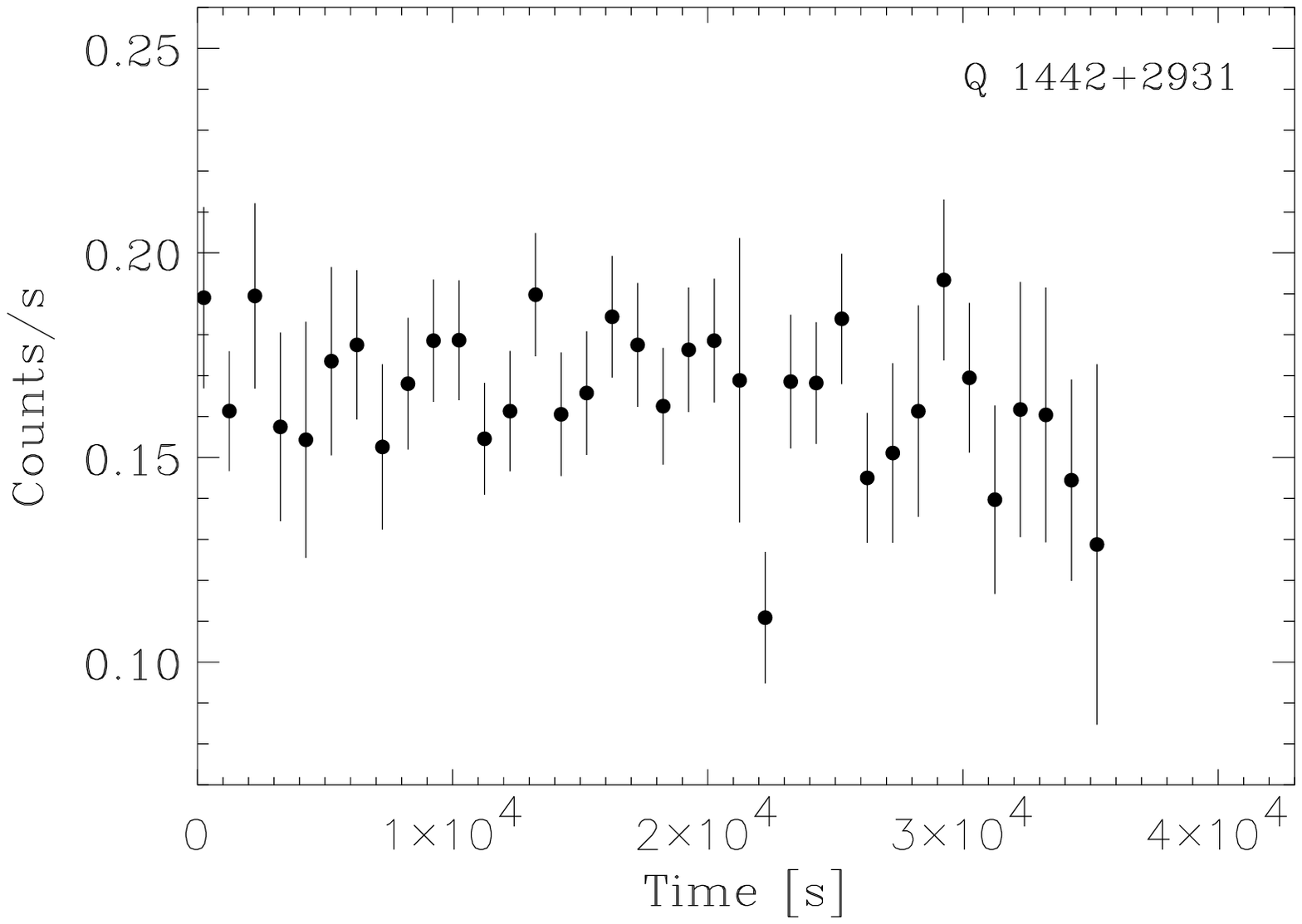,height=4.2truecm,width=8.5truecm,%
 bbllx=76pt,bblly=371pt,bburx=532pt,bbury=693pt,clip=}
\caption[]{ Combined PN+MOS, background subtracted, $0.2-10$ keV lightcurves
of the four quasars. The time binning is 500 sec for the radio-loud
quasars and 1000 sec for the radio-quiet quasars.}
\label{figure:lightcurve}
\end{figure}

\subsection{The light curves}

We calculated the  0.2$-$10 keV light curves for all four objects by 
extracting the photons from a 
circular region centered on the source with a radius of 45\arcsec.
This extraction radius was chosen to avoid contamination
from  nearby  objects and it contains about 90 \% of the 
source photons, 
using the  encircled energy function given by Ghizzardi \& Molendi (2001).
Only single and double events  (i.e. with pattern 0$-$4 for the PN camera
 and 0$-$12 for the MOS cameras) and with quality flag 0 were chosen.
The time bin size was set to be 500 sec for the radio-loud objects
and to 1000 sec for the radio-quiet ones and 
we used only the time range for  which all EPIC cameras were switched on.

The backgrounds were determined  with the same selection criteria 
from source free regions on the same chips  and subtracted from the
source light curves.
The co - added PN and MOS net light curves for the four objects are shown 
in {\bf Fig. \ref{figure:lightcurve}}.
 
The light curve of \X21 is consistent with a constant average
flux during the observation period.
A Kolmogorov-Smirnov test gives a $\leq$ 15\% probability that 
the short time variability is only that expected from purely statistical 
fluctuations. 
However, no definite time scale can be deduced from the
relatively noisy data. 
 
\P2126 shows a slight increase of the count rate  with time of 
$\sim (4.8\pm2.7)\times10^{-6}$ cts~s$^{-2}$ as well as 
short time flux variations  of $\sim$ 5\% on time scales of $\sim$1 ksec.
These variations seem to occur in a well organized linear
fashion but the counting statistics  are insufficient for 
a more detailed analysis.
In both cases these variations are moderate and not unusual amongst 
radio loud quasars.  From the observed variability we can estimate 
 a lower limit to the radiative efficiency 
 $\eta \ma 5\times 10^{-43}\Delta L/\Delta t$ (Fabian 1979).
With the  above given  time scale and the luminosity 
determined from the spectral analysis (see Sec. \ref{xspec}) we
obtain $\eta \ma 8$. This value largely exceeds the theoretical
limit of accretion onto a black hole implying  enhancement of 
the emission by relativistic beaming.

The light curves of the two radio-quiet objects 
\Q0  and  \1442 are  consistent with a constant flux 
and show no statistically significant variations.
The drop of the count rate at the end of the  \1442 observation and the 
one exceptional low point coincide with strong flaring activity of
the background and are thus very likely not source intrinsic
intensity variations.

\section{Spectral analysis}
\label{xspec}

We have performed the spectral analysis of PN, MOS and RGS data 
 for both radio-loud quasars 
using the latest available versions of the response matrices,
released in April 2002 (PN) and in March 2002 (MOS).
The RGS response matrices were created with the SAS package
$rgsrmfgen$.
In the case of the radio-quiet quasars we have analyzed only
the PN and MOS data, because of the low photon statistics in the
RGS spectra.
As a preliminary step, we have created  Good Time Interval  (GTI) 
files to check for time intervals with high background which 
could contaminate significantly the source photons. 
For both the radio-loud sources the background remained low during the 
whole observation, so that no time intervals had to be excluded from 
the subsequent analysis. 
During the  observations of the two radio-quiet objects 
several background flares were found  and 
the flaring time intervals were excluded in the spectral analysis. 

We also checked for pile up exploiting the XMMSAS task $epatplot$, both 
for the PN and the MOS cameras. We found no indications for it 
in the radio-quiet sources and in \X21,
whereas for \P2126 signs for pile up are present for the PN camera 
with  a count rate of $\sim $2.7 cts/s which is close to the  critical value
for a point source in the Extended Full Window mode. 
No pile up was found for the MOS cameras. 

For the PN camera we extracted the photons from a 
circular region of radius 45\arcsec ~ centered on the X-ray positions
of all four quasars. 
This extraction radius corresponds to the maximum allowed to avoid
the chip boundaries. 
In the case of \P2126 we also excluded from the analysis the four 
central (RAW-) pixels in order to avoid pile up.
The same extraction radius as for the PN was used for the MOS cameras.
The backgrounds were extracted from source free regions 
with the same radius from positions near the source.
Only single and double events with flag 0 were selected for the PN, 
whereas  only photons with pattern 0-12 and flag 0
were chosen for the MOS (for details on the XMM detectors 
see Ehle \etal 2001). 
Finally, the produced spectra were binned to contain at least 50 and 30 
photons  
per energy channel for the radio-loud and radio-quiet sources, 
respectively, in order to have  a sufficient signal to noise ratio 
and to  allow
the use of the $\chi^2$ statistics for the fit. 

 To perform the spectral analysis of RGS data we have used the 
standard  science data files created by the RGS Pipeline Processing
binned to contain at least 30 photons per energy channel.

\subsection{\X21}

With  the above selection criteria we obtained a total of $\ma$ 43200 net 
counts from the PN camera for a spectral fit. 
The hardness ratios of the counts in the 0.2$-$1~keV / 2.5$-$10~keV
band show slight variations over the observation, not obviously
correlated with the count rate. We therefore checked the spectral
changes by dividing the observation interval in three parts. 
In all three intervals the fitted  spectral power law 
slopes, assuming galactic $N_{\rm H}$ or fitting the absorbing column density,
  remained the same inside the statistical uncertainties. 
The differences were marginal and therefore
 we combined the whole data set for the spectral analysis. 

\begin{figure}
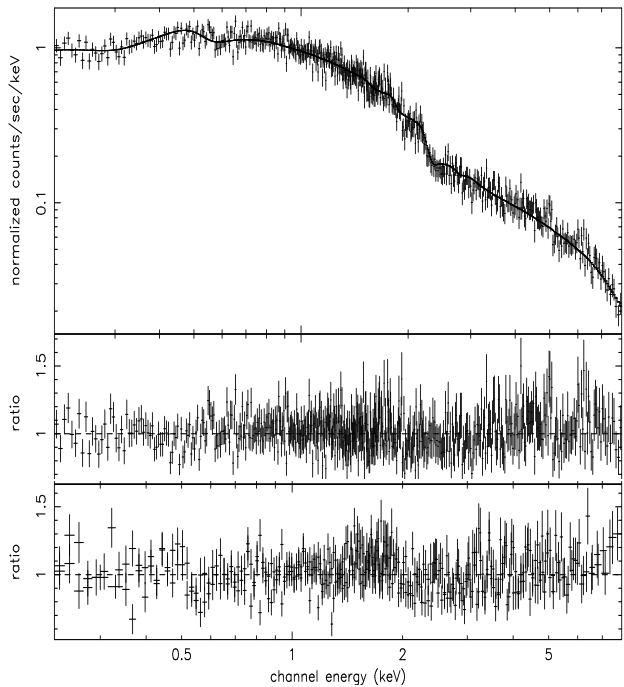

\psfig{figure=powl.ps,height=6.3truecm,width=8.3truecm,angle=-90,%
 bbllx=113pt,bblly=43pt,bburx=505pt,bbury=711pt,clip=}
\psfig{figure=mos_powl5.ps,height=2.7truecm,width=8.3truecm,angle=-90,%
 bbllx=383pt,bblly=43pt,bburx=554pt,bbury=711pt,clip=}
\caption[]{Power law fit with free absorption for \X21 in the $0.2 - 8$
~keV energy range. The upper panel shows the PN fit, the middle panel
 the ratio between data and model; the lower panel gives for a comparison
the ratios for the power law fit to the combined MOS1 plus MOS2 data.}
\label{figure:powl}
\end{figure}

We first fitted the PN data  (see Table \ref{fits}) 
with a simple power law model with free absorption 
for the  $0.2-8$ keV energy range, leaving out the inherently noisy data
above 8 keV. The fit and the resulting residuals, as ratio between model and
data, are given in Fig.~\ref{figure:powl}. 
This fit yielded a photon index $\Gamma = 1.53 \pm 0.02$ and
$N_{\rm H}=(2.94\pm 0.32)\times 10^{20}$ cm$^{-2}$, slightly in excess 
of the galactic
value  of $N_{\rm H}^{\rm gal}=2.10\times 10^{20}$ cm$^{-2}$.
The fit is acceptable with a reduced $\chi^{2}_{\rm red} = 1.08$/528 d.o.f.
A similar fit with $N_{\rm H}$ fixed to the galactic value is
worse with  $\Gamma = 1.49 \pm 0.01$ and a 
reduced $\chi^2 = 1.11$.
We further tried a broken power law model  and obtained
a very flat slope ($\Gamma _{\rm soft} = 1.03$) at
low energies, a slope of $\Gamma _{\rm hard}=1.50$ at high energies, 
similar to the value of the single power law fit, 
 a break energy  of $E_{\rm break}=0.7$ keV, but with an  $N_{\rm H}$ lower 
than the galactic value, with a reduced $\chi^{2}_{\rm red} = 1.04/526$. 
Fixing the absorption at the 
galactic value yields an  equally acceptable fit    
($\chi^{2}_{\rm red} = 1.05/527$) with
nearly identical parameters, except for a  steeper slope
($\Gamma _{\rm soft} = 1.27$) at low energies.
 An F-test gives an improvement for the 
broken power law fit at only the $\sim $86\% confidence
level, therefore we will from
now on only consider single power law fits.

The power law fit to the MOS data gave slightly different results, summarized
in Table \ref{fits}. 
As  can be seen, the reduced $\chi^{2}$ are larger
for MOS2 and for both MOS1 and MOS2 the fitted slopes are 
considerably flatter than those for the PN, more than usually found 
when comparing MOS and PN data. Deviations from a simple power law spectrum
could be responsible for the observed differences.
For the MOS2 we also get a column density lower than the galactic value.

\setcounter{table}{1}
\begin{table} 
\small
\tabcolsep1ex
\caption{\label{fits} Results for power law fits for \X21 in the energy band $0.2-8.0$ keV.}
\begin{tabular}{lcccc}
\noalign{\smallskip} \hline \noalign{\smallskip}
\multicolumn{1}{c}{Detector} &
\multicolumn{1}{c}{$N_{\rm H}$} & \multicolumn{1}{c}{$\Gamma $} &
\multicolumn{1}{c}{Normalization} & \multicolumn{1}{c}{$\chi^2_{\rm red}$/dof} \\
\multicolumn{1}{c}{  } & \multicolumn{1}{c}{(1)}
&\multicolumn{1}{c}{ } & 
\multicolumn{1}{c}{ (2) } & \multicolumn{1}{c}{}\\
\noalign{\smallskip} \hline \noalign{\smallskip}
 PN & 2.94$\pm0.32$ &
   1.53$\pm$0.02 & 0.87$\pm$0.15 & 1.08/528\\
 PN & galactic &
   1.49$\pm$0.01 & 0.83$\pm$0.07 & 1.11/529\\
  MOS1 & 2.34$\pm0.80$ &
   1.27$\pm$0.03 & 0.95$\pm$0.03 & 1.08/188\\
 MOS1 & galactic &
   1.27$\pm$0.02 & 0.94$\pm$0.02 & 1.07/189\\
   MOS2 & 0.55$\pm0.41$ &
   1.27$\pm$0.03 & 0.91$\pm$0.03 &1.28/189\\
 MOS2 & galactic &
   1.33$\pm$0.02 & 0.97$\pm$0.02 & 1.34/190\\
 MOS1+MOS2 & 1.36$\pm0.52$ &
   1.27$\pm$0.03 & (a) & 1.23/383\\
 MOS1+MOS2 & galactic &
   1.30$\pm$0.02 & (a) & 1.24/384\\
\noalign{\smallskip}\hline
\end{tabular}
\medskip

 The errors given are at the 90\% level.\\
(1): in units of $10^{20}$ cm$^{-2}$;
 $N_{\rm H}^{\rm gal}=2.10\times 10^{20}$ cm$^{-2}$. \\
(2): Normalization at 1 keV in 10$^{-3}$ ph/keV/cm$^2$/s. \\
(a): Different normalizations for the individual detectors.\\
\end{table}

To improve the statistics, we combined  the MOS1 and MOS2 data.
The power law fits yielded  similar results as above, with slopes 
 still flatter than obtained for the PN (see Fig. ~\ref{figure:powl}).
Interestingly, the slopes found for the simple power law fit to the MOS
 data are very similar to the values  of the low-energy slopes of the 
broken power law fit to the PN data (with galactic $N_{\rm H}$). 

However, the quality of the MOS data appears to be lower,
the residuals are much noisier and the obtained $\chi^2$s are 
 worse than those  from the fits to the PN data, which can be
seen in Fig.  \ref{figure:powl} where the lower panel shows the
ratios for the MOS fit.
Whether remaining calibration uncertainties or an intrinsic
source  spectrum  which cannot adequately be described by a simple
power law,  account for the observed discrepancies cannot be 
distinguished with the current data.

We further fitted the RGS data with an absorbed power law
in the  $0.35-2$ keV energy range.
 The fits to the RGS1 and RGS2 data yielded different results,
with power law slopes $\Gamma =1.16\pm 0.18$ 
($\chi^2_{\rm red}= 1.18/37$ d.o.f) and
$\Gamma =1.64\pm 0.38$ ($\chi^2_{\rm red}= 0.91/43$ d.o.f)
assuming galactic absorption for RGS1 and RGS2, 
respectively. The fits with free absorption gave similar slopes
 with $N_{\rm H}$ consistent with the galactic value
for the  RGS1 whereas in the case of the RGS2
 a flatter slope ($\Gamma = 1.52\pm0.50$)
with $N_{\rm H}$ lower than the galactic value is found.
Due to the discrepancies between the two instruments, and
to improve the photon statistics, we combined the RGS1 and RGS2 
data, tolerating some data degradation and inferior resolution
compared to the separate fits. 
The results yielded flat slopes, in between those from the analogous fits
to the MOS and PN data ($\Gamma = 1.41\pm0.13$,
 $\chi^2_{\rm red}= 1.14$/81 d.o.f. for galactic $N_{\rm H}$;
 $\Gamma = 1.26\pm0.32$, $\chi^2_{\rm red}= 1.15$/80 d.o.f
with an $N_{\rm H}$ value compatible with zero).
There are no obvious structures to be seen in the residuals, but the
quality of the RGS data is rather low.
Considering the discrepancies among the various
instruments, we did not attempt to combine and fit the
PN and MOS data or the PN and RGS data together, 
but rely in the following mainly on the PN data.

We tried  other models like a thermal
bremsstrahlung and a constant density ionized disk model
(Ballantyne  \etal 2001).
The thermal bremsstrahlung model yielded a temperature of 
$kT=9.8$ keV;  however, the fit is unacceptable with a 
reduced $\chi ^{2}_{\rm red } = 1.42/529$ d.o.f.
The ionized disk model (assuming galactic $N_{\rm H}$)  resulted in an
ionization parameter $\xi = 1.024$, a photon index of the incident power law
$\Gamma = 1.5$, a reflection fraction $R=1.73\times 10^{-7}$, 
and a reduced $\chi ^{2}_{\rm red } = 1.13/527$ d.o.f.
Thus, the ionized disk model provides only a poor fit,
the reflection fraction is very small, and the obtained slope is
very similar to that of the power law fit.

Single power law fits to the data in the hard energy band ($\ma 2$ keV)
result in excellent fits with slightly flatter slopes (see Table \ref{hard}).
 An extrapolation
of those power laws to lower energies seem to indicate a deficit of 
flux below $\sim$ 1 keV. However, small changes in the energy range 
used for the fit (especially when the inherently noisy data $\ma$ 8 keV
are left out from the fit)  result in acceptable fits ($\chi^2_{\rm red}
\sim 1$) with  slightly differing power law slopes  and fitted
 $N_{\rm H}$ values in accordance to the galactic absorption.
 
The addition of an extra emission component to the power law,
for example a black body with $kT = 0.25\pm0.02$ keV, contributing about
5$-$8\% to the flux at 1~keV, provides an acceptable fit
over the whole 0.2$-$10.0 keV energy band  ($\chi^2_{\rm red}
 = 1.036$/528 d.o.f.) with a power law slope  of
$\Gamma = 1.43\pm0.02$ and fixed galactic absorption. 
 
Finally, we tried a power law with fixed galactic absorption over the whole
energy band, allowing for extra absorption at the redshift of the source.
The fit is acceptable ($\chi^2_{\rm red}=1.065/528$ d.o.f.) 
with  $\Gamma =1.53\pm 0.02$
and an intrinsic column density of 
$N_{\rm H,z}=(1.09\pm0.04)\times 10^{21}$ cm$^{-2}$.
A similar fit could be achieved with galactic absorption plus an intrinsic
warm absorber ($absori$ model in XSPEC).
The fit is acceptable ($\chi^2_{\rm red}$=1.067/527 d.o.f) with
 $\Gamma = 1.53\pm0.03$. The column density of the warm
absorber $N_{\rm warm} = (1.31^{+>3}_{-0.9})\times10^{21}$ cm$^{-2}$
is rather ill determined.

To investigate the presence of a Gaussian line around $\sim $5 keV
($\sim $17 keV in the quasar rest frame) claimed
by Yaqoob \etal (1999) from an ASCA observation,
 we restricted our analysis to the hard energy band (2-8 keV).
We fitted a power law with galactic absorption
and an additional narrow ($\sigma = 0.01$ keV) Gaussian line.
The result for the fit is given in Table \ref{hard}.
The equivalent width of this narrow line would be about  12 eV 
in the quasar's rest frame,
much below the $\sim 300$ eV claimed by Yaqoob \etal (1999).
Leaving the width of the line free in the fit resulted as well in
a narrow line with nearly identical parameters and large errors.
We further tried a power law fit with inclusion of
an absorption edge at $\sim $5 keV and  obtained an equally
acceptable fit (see Table \ref{hard}).

\setcounter{table}{2}
\begin{table} 
\small
\tabcolsep1ex
\caption{\label{hard} PN spectral fits for \X21 in the hard energy band $2-8$ keV and galactic absorption.}
\begin{tabular}{lcccc}
\noalign{\smallskip} \hline \noalign{\smallskip}
\multicolumn{1}{c}{Model} & \multicolumn{1}{c}{$\Gamma $} &
\multicolumn{1}{c}{$E$} &
\multicolumn{1}{c}{$\sigma $/$\tau $} &
 \multicolumn{1}{c}{$\chi^2_{\rm red}$/dof} \\
\multicolumn{1}{c}{  } & \multicolumn{1}{c}{ } & \multicolumn{1}{c}{(1)} &
\multicolumn{1}{c}{(2) } &
 \multicolumn{1}{c}{}\\
\noalign{\smallskip} \hline \noalign{\smallskip}
pow & 1.44$\pm$0.04 &  &   & 0.87/196\\
pow + gauss &
   1.44$\pm$0.05 & 4.96$\pm$0.06 &  (0.01) & 0.84/194\\
pow + edge &  1.40$\pm$0.07 &  5.10$\pm$0.20 &  0.10$\pm$0.10 & 0.85/194\\
\noalign{\smallskip}\hline
\end{tabular}
\medskip

 The errors given are at the 90\% level.\\
(1) Line energy in keV; ~ the normalization of the line  is\\ 
     3.02$\pm$2.57$\times $10$^{-6}$ ph/keV/cm$^2$/s.\\
(2) Line width ($\sigma $) or depth of the absorption edge ($\tau $) in keV.\\
\end{table}

Both, the fit with a Gaussian line and an absorption edge,
 improve the $\chi^2_{\rm red}$ only marginally, 
 the normalizations of the models are small and
not well constrained and the improvements are
statistically not significant.
We also tried to add a Gaussian line to the model of a power law plus 
black body and galactic absorption, resulting in  $kT = 0.25$ keV,
 $\Gamma =1.43$, a narrow line at $E=4.95$ keV  
and 
$\chi^2_{\rm red}=1.03/525$ d.o.f.
Also in this case  the introduction of the line
does not improve the fit significantly.
  
 From the simple power law fit with free $N_{\rm H}$ in the whole energy
band we obtain a 2-10 keV flux and luminosity of
$f_{\rm 2-10~keV}=4.65\times 10^{-12}$ \ergs and
 $L_{\rm 2-10~keV}=1.03\times 10^{47}$ erg s$^{-1}$, respectively.
For the ROSAT band we get $f_{\rm 0.1-2.4~keV}=3.47\times 10^{-12}$ \ergs
and $L_{\rm 0.1-2.4~keV}=7.73\times 10^{46}$ erg s$^{-1}$.

\subsection{\P2126}

Applying the same selection criteria as for \X21, we collected
a total of $\ma$ 44300 net counts from the PN camera. We started fitting
a simple power law model in the $0.2-10.0$ keV range and 
obtained a photon index $\Gamma =1.51\pm 0.02$, an
$N_{\rm H}=1.1\times 10^{21}$ cm$^{-2}$, in excess of the galactic value and
a $\chi^2_{\rm red}=0.98/492$ d.o.f.
(see Table \ref{fits2}). 
A single power law fit with fixed galactic absorption
is not acceptable with  $\chi^2_{\rm red}=1.85/493$ d.o.f. 
  
\begin{figure}
\psfig{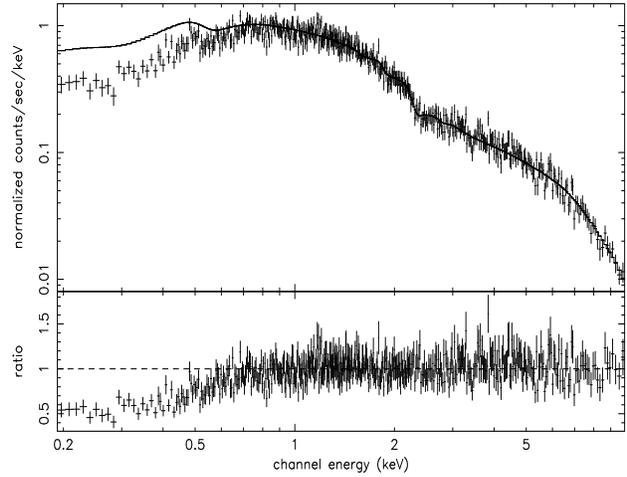}
\caption[]{Power law fit with galactic absorption for \P2126 over 
 the restricted energy band 2$-$10 keV, extrapolated to lower energies.}
\label{figure:plext} 
\end{figure}

A power law fit, limited to the hard energy band ($2.0-10.0$ keV),
with absorption fixed to the galactic value is acceptable
($\chi^2_{\rm red}=0.88/197$ d.o.f.) with  $\Gamma =1.43\pm 0.04$.
The extrapolation of this power law fit to lower energies
(see Fig. ~\ref{figure:plext}) clearly demonstrates the necessity for 
more absorption at low energies.
  
The fit further indicates that, although excellent power law fits 
can be achieved, there might be spectral deviations from this 
simple model. 
As an indication for this we notice, that   
the same fit, limited to the hard energy band ($3.0-10.0$ keV),
with fixed galactic absorption is acceptable 
($\chi^2_{\rm red}=0.99/133$ d.o.f.) with a  different slope
of  $\Gamma =1.50\pm 0.07$.
  
\begin{figure}
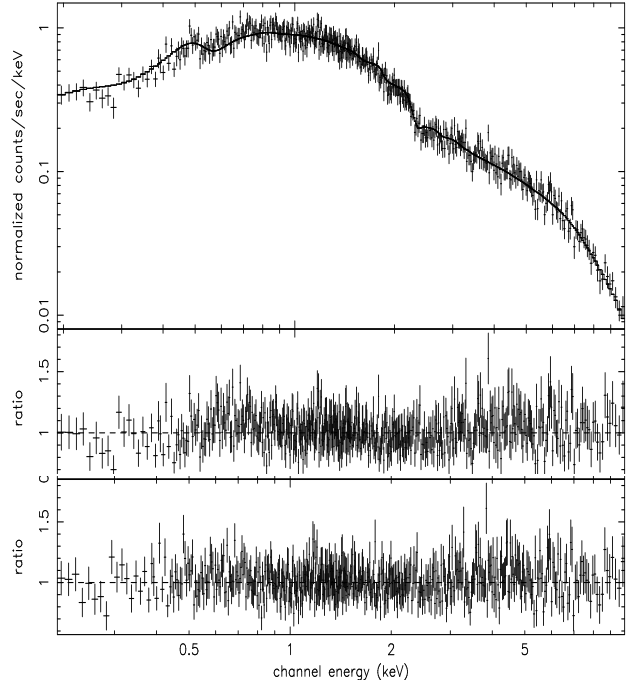

\psfig{figure=2126powl.ps,height=6.3truecm,width=8.3truecm,angle=-90,%
 bbllx=113pt,bblly=43pt,bburx=505pt,bbury=711pt,clip=}
\psfig{figure=absori.ps,height=2.7truecm,width=8.3truecm,angle=-90,%
 bbllx=383pt,bblly=43pt,bburx=554pt,bbury=711pt,clip=}
\caption[]{Power law fit with free absorption to the PN data for \P2126
(upper and middle panel). The lower panel shows the ratios between
data and model for the warm absorber fit with galactic absorption
to the PN data.}
\label{figure:2126pow}
\end{figure}

Fitting a power law over the whole energy band, fixing the galactic absorption 
and allowing for extra absorption at the redshift of the source,
gives a good fit ($\chi^2_{\rm red}=0.90/492$ d.o.f.) with extra
$N_{\rm H,z}=1.40\times 10^{22}$ cm$^{-2}$ and $\Gamma =1.47\pm 0.02$. 
Allowing the redshift of the absorber to vary results in a value
in agreement with the redshift of the quasar.
We checked if a warm absorber could be responsible for the
extra absorption, performing a fit with the $absori$ model.
The ratios between the data and the model of the resulting fit are
shown in the lower panel of Fig. \ref{figure:2126pow} with extra 
$N_{\rm H,z}=1.35\times 10^{22}$ cm$^{-2}$ at the redshift of the
source,  $\Gamma =1.47\pm 0.02$, the absorber temperature fixed
at $T=3\times 10^{4}$ K and $\chi^2_{\rm red}=0.90/490$ d.o.f.
However, the ionization parameter is not well
constrained and the fit is statistically equivalent to the
one with a neutral absorber at the source's redshift. 

\setcounter{table}{3}
\begin{table} 
\small
\tabcolsep1ex
\caption{\label{fits2} Results for power law fits for \P2126.}
\begin{tabular}{lcccc}
\noalign{\smallskip} \hline \noalign{\smallskip}
\multicolumn{1}{c}{Detector} &
\multicolumn{1}{c}{$N_{\rm H}$} & \multicolumn{1}{c}{$\Gamma $} &
\multicolumn{1}{c}{Normalization} & \multicolumn{1}{c}{$\chi^2_{\rm red}$/dof} \\
\multicolumn{1}{c}{  } & \multicolumn{1}{c}{(1)}
&\multicolumn{1}{c}{ } & 
\multicolumn{1}{c}{ (2) } & \multicolumn{1}{c}{}\\
\noalign{\smallskip} \hline \noalign{\smallskip}
 PN$^{(a)}$ & {\bf 1.12$\pm0.06$} &
   1.51$\pm$0.02 & 0.99$\pm$0.02 & 0.98/492\\
 PN$^{(a)}$ & galactic &
   1.30$\pm$0.01 & 0.77$\pm$0.01 & 1.85/493\\
 PN$^{(b)}$ & {\bf 0.49$\pm8.28$} &
   1.50$\pm$0.20 & 1.00$\pm$0.49 & 1.00/132\\
 PN$^{(b)}$ & galactic &
   1.50$\pm$0.07 & 1.00$\pm$0.12 & 0.99/133\\
 PN$^{(c)}$ & {\bf 3.10$\pm1.99$} &
   1.49$\pm$0.41 & 0.87$\pm$0.33 & 0.87/196\\
 PN$^{(c)}$ & galactic &
   1.43$\pm$0.04 & 0.88$\pm$0.04 & 0.88/197\\
  MOS1$^{(a)}$ & {\bf 1.29$\pm0.09$} &
   1.35$\pm$0.03 & 1.74$\pm$0.06 & 1.26/229\\
 MOS1$^{(a)}$ & galactic &
   1.11$\pm$0.01 & 1.31$\pm$0.02 & 2.36/230\\
   MOS2$^{(a)}$ & {\bf 1.13$\pm0.09$} &
   1.34$\pm$0.03 & 1.71$\pm$0.06 &1.22/233\\
 MOS2$^{(a)}$ & galactic &
   1.14$\pm$0.02 & 1.36$\pm$0.02 & 1.90/234\\
\noalign{\smallskip}\hline
\end{tabular}
\medskip

 The errors given are at the 90\% level.\\
(1):  in units of  $10^{21}$ cm$^{-2}$;
 $N_{\rm H}^{\rm gal}=4.90\times 10^{20}$ cm$^{-2}$. \\
(2): Normalization at 1 keV in 10$^{-3}$ ph/keV/cm$^2$/s. \\
(a): Energy band $0.2-10.0$ keV.\\
(b): Energy band $3.0-10.0$ keV.\\
(c): Energy band $2.0-10.0$ keV. \\
\end{table}

The results of the power law fits for the MOS data are shown
in Table \ref{fits2}. 
As in the case of \X21 the fits are systematically worse than
those for the PN, with significantly flatter slopes 
and larger $\chi^2$s. Apart from these differences, the MOS
data show evidence for excess absorption in \P2126 as well.
A simple power law fit with excess absorption doesn't
provide a good description of the spectrum and 
the remaining residuals indicate 
a more complicated structure, but the photon statistics
of the MOS data  are insufficient for a more detailed modeling.

Fitting the same models to the RGS data gives results consistent 
with those obtained for the PN, but with flatter slopes
and lower excess absorption  
 ($\Gamma \sim 1.2, ~ N_{\rm H}=8.7\times 10^{20}$cm$^{-2}$  for RGS1, 
$\Gamma \sim 1.4, ~N_{\rm H}=8.3\times 10^{20}$cm$^{-2}$ for RGS2,
$\Gamma \sim 1.2,  ~N_{\rm H}=6.6\times 10^{20}$cm$^{-2}$ 
for the combined fit). 
The residuals show an excess of emission around $1.7$ keV
which we further tried to model with a Gaussian line with fixed 
energy ($E=1.72$ keV) and width ($\sigma =0.05$ keV),
with equivalent widths of $\sim$ 129 eV for the RGS1 and $\sim$
67 eV for the RGS2. 
This allowed  us to recover values of the parameters similar
 to the PN case both
for the absorption ($N_{\rm H}=1.0\times 10^{21}$cm$^{-2}$) and
the power law slope ($\Gamma =1.55\pm 0.41$).
However, the rest frame energy of the line would be
$\sim 7.3$ keV, not immediately recognizable as any known feature.

We tried additional models relying again on the PN data alone.
An ionized disk model yielded
acceptable results with the following best fit parameters:
 $N_{\rm H}=1.2\times 10^{21}$cm$^{-2}$, ionization parameter $\xi =1.044$,
$\Gamma =1.56$, reflection fraction $R=0.11$,
redshift $z=3.262$ (consistent with the source's redshift)
and $\chi^2_{\rm red}=0.97/489$ d.o.f. 
But the errors on the parameters are large  so that this
fit is not reliable.

A thermal bremsstrahlung model with free $N_{\rm H}$ gives a slightly worse
fit than the power law ($\chi^2_{\rm red}=1.11/491$ d.o.f.),
whereas a broken power law fit results in very low values
of the soft photon index ($\Gamma _{\rm soft} =0.09$ for free  $N_{\rm H}$)
and huge errors.

 Using the results from the simple power law fit with free absorption
in the whole energy band we obtain the following fluxes and 
luminosities: $f_{\rm 2-10~keV}=5.46\times 10^{-12}$ \ergs ,
$L_{\rm 2-10~keV}=2.23\times 10^{47}$ erg s$^{-1}$,
$f_{\rm 0.1-2.4~keV}=3.92\times 10^{-12}$ \ergs and
$L_{\rm 0.1-2.4~keV}=1.61\times 10^{47}$ erg s$^{-1}$.

\subsection{\Q0}

\begin{figure}
\psfig{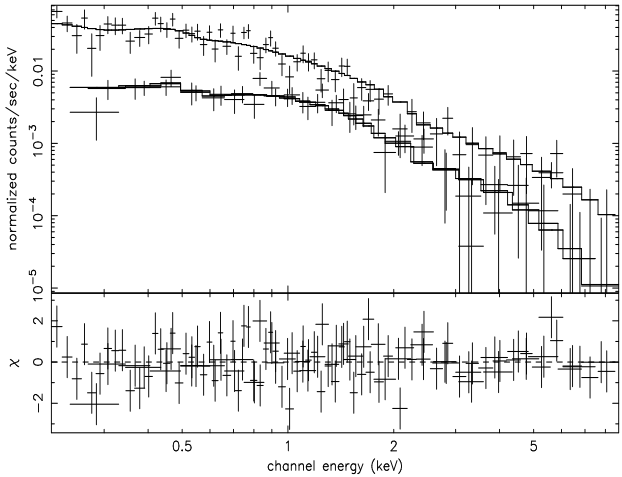}
\caption[]{Combined PN+MOS power law fit with galactic absorption 
for \Q0 in the $0.2 - 9$
~keV energy range. The lower panel gives the
$\Delta \chi^2$ per channel.}
\label{figure:powl0000}
\end{figure}

From the $\ma $ 40 ksec observation 
we collected only $\ma$ 1400 net source counts from the PN camera. 
We tried  simple power law fits in the whole energy band 
for the PN and MOS cameras separately.
In all three cases the fitted $N_{\rm H}$ was consistent with the galactic
value, so we fixed it to reduce the number of fit parameters.
The resulting photon indices are given in Table \ref{Q0fits}.
As they are  consistent with each other within the errors, we combined
the PN and MOS spectra to improve the statistics. A simple power law
fit with galactic absorption yielded a photon index  $\Gamma =2.10\pm 0.11$
and $\chi^2_{\rm red}=0.85/110$ d.o.f. (see Fig. \ref{figure:powl0000}).
No Fe line (expected at $\sim 1.2$ keV in the observer's frame) is  required
by the data. Thus, a simple power law with galactic absorption seems to be
a sufficient model to describe the data.
 Some deviations are visible in the residuals at $\sim 0.9$ keV, which
we tried to model with a broad Gaussian line, however the line's
parameters are not well constrained. The photon counts are insufficient
to model this feature.

With the above parameters for the PN fits we obtain a 2$-$10 keV flux of
$f_{\rm 2-10~keV}=2.55\times 10^{-14}$ \ergs   which results in a  luminosity of $L_{\rm 2-10~keV}=4.82\times 10^{45}$ erg s$^{-1}$   
 of the source. In the ROSAT band we get  
$f_{\rm 0.1-2.4~keV}=7.76\times 10^{-14}$ \ergs and 
$L_{\rm 0.1-2.4~keV}=1.46\times 10^{46}$ erg s$^{-1}$, consistent with the 
values found by Bechtold et al. (1994a) and Kaspi et al. (2000) within
the errors.

\setcounter{table}{4}
\begin{table} 
\small
\tabcolsep1ex
\caption{\label{Q0fits} Results for power law fits and galactic
absorption ($N_{\rm H}^{\rm gal}=1.67\times 10^{20}$ cm$^{-2}$) for \Q0 
in the energy band $0.2-9.0$ keV.}
\begin{tabular}{lccc}
\noalign{\smallskip} \hline \noalign{\smallskip}
\multicolumn{1}{c}{Detector} & \multicolumn{1}{c}{$\Gamma $} &
\multicolumn{1}{c}{Normalization} & \multicolumn{1}{c}{$\chi^2_{\rm red}$/dof} \\
\multicolumn{1}{c}{  } & \multicolumn{1}{c}{ } & 
\multicolumn{1}{c}{ (1) } & \multicolumn{1}{c}{}\\
\noalign{\smallskip} \hline \noalign{\smallskip}
 PN &
   2.19$\pm$0.12 & 1.31$\pm$0.14 & 0.91/64\\
  MOS1 &
   1.98$\pm$0.40 & 1.58$\pm$0.19 & 0.67/21\\
   MOS2 &
   1.82$\pm$0.23 & 1.46$\pm$0.23 &0.52/23\\
 PN+MOS &
   2.10$\pm$0.11 & (a) & 0.85/110\\
\noalign{\smallskip}\hline
\end{tabular}
\medskip

 The errors given are at the 90\% level.\\
(1): Normalization at 1 keV in 10$^{-5}$ ph/keV/cm$^2$/s. \\
(a): Different normalizations for the individual detectors.\\
\end{table}

\subsection{\1442}

A total of $\ma $ 3800 net counts were collected for the source in the
PN camera. Again, we first tried simple power
law fits for the PN and MOS instruments separately, which resulted 
in absorption column densities 
consistent with the galactic value. We therefore fixed 
the $N_{\rm H}$ to the galactic value  and  obtained the results 
given in Table \ref{Q14fits}.

\setcounter{table}{5}
\begin{table} 
\small
\tabcolsep1ex
\caption{\label{Q14fits} Results for power law fits and galactic
absorption ($N_{\rm H}^{\rm gal}=1.56\times 10^{20}$ cm$^{-2}$) for \1442 
in the energy band $0.2-9.0$ keV.}
\begin{tabular}{lccc}
\noalign{\smallskip} \hline \noalign{\smallskip}
\multicolumn{1}{c}{Detector} & \multicolumn{1}{c}{$\Gamma $} &
\multicolumn{1}{c}{Normalization} & \multicolumn{1}{c}{$\chi^2_{\rm red}$/dof} \\
\multicolumn{1}{c}{  } & \multicolumn{1}{c}{ } & 
\multicolumn{1}{c}{ (1) } & \multicolumn{1}{c}{}\\
\noalign{\smallskip} \hline \noalign{\smallskip}
 PN &
   1.92$\pm$0.07 & 4.31$\pm$0.30 & 0.87/105\\
  MOS1 &
   1.82$\pm$0.11 & 5.19$\pm$0.48 & 2.03/48\\
   MOS2 &
   1.76$\pm$0.13 & 5.15$\pm$0.51 &0.84/48\\
 PN+MOS &
   1.87$\pm$0.05 & (a) & 1.14/203\\
\noalign{\smallskip}\hline
\end{tabular}
\medskip

 The errors given are at the 90\% level.\\
(1): Normalization at 1 keV in 10$^{-5}$ ph/keV/cm$^2$/s. \\
(a): Different normalizations for the individual detectors.\\
\end{table}

\begin{figure}
\psfig{figure=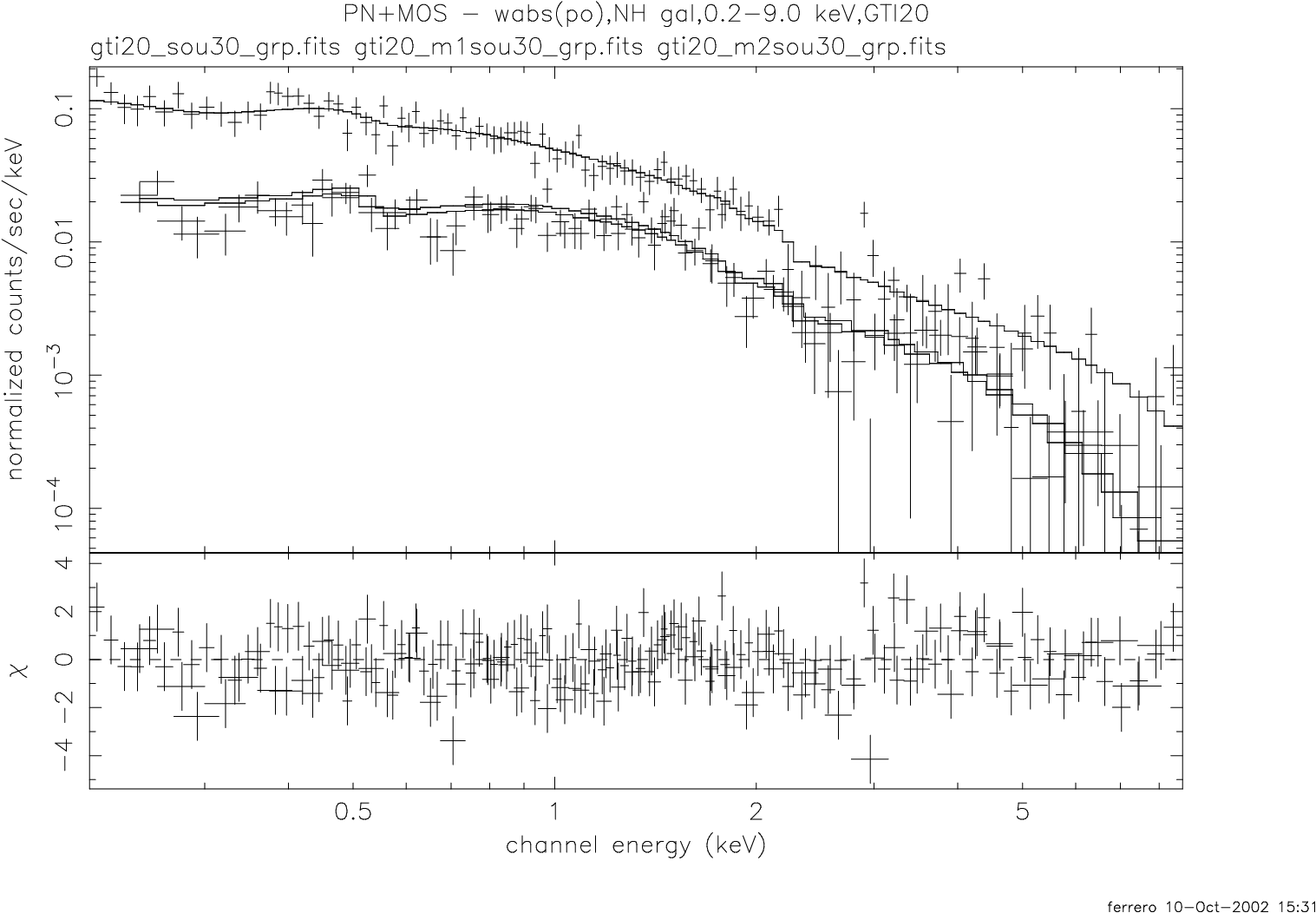,height=6.3truecm,width=8.3truecm,angle=-90,%
bbllx=113pt,bblly=43pt,bburx=557pt,bbury=711pt,clip=}
\caption[]{Combined PN+MOS power law fit with galactic absorption 
for \1442 in the $0.2 - 9$
~keV energy range. The lower panel gives the $\Delta \chi^2$ per channel.}
\label{figure:powl1442}
\end{figure}

The fit to the very noisy MOS1 data gave a high $\chi^2_{\rm red}$, in contrast to the 
fit for the MOS2 camera. The MOS1 data could also be responsible for the high
$\chi^2_{\rm red}$ of the combined  PN + MOS fit 
 (see Fig. \ref{figure:powl1442}). All fits indicate 
that a power law  describes the spectrum of the source quite well
and that no excess absorption is required. The MOS data, but not the PN data,
 seem to require an extra emission component at $\sim 1.5$ keV, 
which we tried to model as a broad Gaussian line.   
From the fit of the co-added MOS1 and MOS2 data we obtained a line
energy $E=1.48\pm 0.12$ keV; however the line width is not constrained
and the fit doesn't improve significantly after the addition of the line.
The low quality of the MOS data and the small number of counts argue against
the real presence of this feature. The addition of the PN data leads to 
a shift of the free fitted line center to an energy
below 0.2 keV, with large errors on both the
energy of the line and its width.
We note that a  redshifted  Fe line would be expected at
$\sim 1.7$ keV, where no feature is observed.
   
The 2$-$10 keV flux and luminosity resulting from the PN fit are
 $f_{\rm 2-10~keV}=1.25\times 10^{-13}$ \ergs and
 $L_{\rm 2-10~keV}=5.82\times 10^{45}$ erg s$^{-1}$, respectively. 
For the ROSAT band
they are  $f_{\rm 0.1-2.4~keV}=2.08\times 10^{-13}$ \ergs and
$L_{\rm 0.1-2.4~keV}=9.64\times 10^{45}$ erg s$^{-1}$, consistent
inside the errors with no flux variations between the ROSAT 
(Reimers et al. 1995)
and XMM
observations ten years apart.

\section{Discussion}
 
We have presented the results of detailed spectral analysis of two high$-$z,
very luminous, radio-loud quasars and two high$-$z radio-quiet quasars.
XMM's high sensitivity and wide energy band allowed a reliable determination of 
the quasars' spectra over a source intrinsic energy band up to $\sim$ 50 keV
and to address the question of intrinsic absorption in the objects.

\subsection{Radio-loud objects}
 
A simple power law  with a slope of  $\Gamma \sim 1.53$ provides a 
reasonably good fit to the data from PKS~2149$-$306,
with absorption near the galactic value. However, the fact that MOS and PN
data yield different slopes, that fits over restricted energy ranges 
for the same instrument yield slightly different power law
indices, and that the residuals show some `systematic'
variations indicates, that a simple power law is perhaps not the best
 description for the spectrum.   
But the deviations from a straight power law are small and statistically
only separable by increasing the signal to noise ratio in a longer
observation.
 
Depending on the models fitted to the data the results for the absorption
column density change slightly. Adding a small black body component
with $kT\sim 0.25$ keV, which might be a reflection component in the 
rest frame of the source, to the power law gives an acceptable fit with
 galactic absorption. 
Another interesting possibility could be that we are seeing the soft
X-ray bump, never observed up to now, produced in blazars
by the Comptonization of external UV radiation by electrons in the jet
(Sikora et al. 1997).

Source intrinsic, extra  warm or cold absorption models predict
a column density
of $\sim 10^{21}$ cm$^{-2}$ in the rest frame of the source.
Considering the quality of the available data we conclude, however, that
there is no strong evidence for substantial absorption in excess of
the galactic value towards PKS~2149$-$306.

The  source was previously observed by ROSAT and ASCA (Siebert \etal 1996,
Cappi \etal 1997), BeppoSAX (Elvis \etal 2000), and Chandra (Fang \etal 2001). 
A comparison between the different instruments indicates spectral and
flux variability. The ASCA power law slopes are comparable
to ours and  the spectra show indications of excess absorption of a
few times 10$^{20}$ cm$^{-2}$. BeppoSAX claims a broken power law with
a similar flat slope and soft excess emission below $\sim$ 0.8 keV,
   while the {\it Chandra} spectra are in agreement
with galactic absorption, and a very flat power law slope 
($\Gamma \sim 1.26$). 
The 2$-$10 keV flux measured in October 1994 
by ASCA is about 30\% higher than that obtained  by {\it Chandra}, BeppoSAX (in 1997),
and XMM ($\sim 7 - 8\times10^{-12}$ \ergs ).  
The RASS flux in 1990/91 seems to have been a factor of two lower
(Schartel \etal 1996). However, the rather uncertain spectral slope
makes an exact comparison problematic.
  
In none of the recent observations (BeppoSAX, {\it Chandra}, and XMM-Newton)
the claimed detection of a line at $\sim$ 5 keV could be confirmed,
nor is there evidence for line emission at other energies.
 
PKS 2126$-$158 clearly shows the presence of extra absorption of the order
of $\Delta {N}_{\rm H} \ma 6\times 10^{20}$ cm$^{-2}$.
At the redshift of the source this corresponds to 
$N_{\rm H,z} \sim 1.4\times10^{22}$ cm$^{-2}$  and a power law
model with galactic absorption plus an intrinsic cold or a
warm absorber, results in excellent fits.  

The object has been observed four times  between 1991-1993 with the 
ROSAT PSPC (Elvis \etal 1994b, Cappi \etal 1997) and in 1993 by ASCA
(Cappi \etal 1997). For all observations the fitted power law slopes 
and the values
of excess absorption are consistent with our results inside the errors
with no indications of temporal variations.
The measured fluxes, even from the early {\it Einstein} IPC observation 
(Worrall \& Wilkes 1990) are identical  inside the statistical 
uncertainties, $f_{2-10~keV} \sim 1\times 10^{-11}$ \ergs.
 The constancy of this flux level is relatively unusual for such a
bright radio-loud quasar.

\subsubsection{X-ray and broad band spectral properties}
\label{411}

The power law fits with free absorption in the whole energy band
yielded for both quasars a slope $\Gamma \la 1.5$.
 This value is typical of
those from {\it Einstein} observations (Wilkes \& Elvis 1987)
but is at the lower limit of the range commonly  found 
for radio-loud quasars from ROSAT and ASCA observations
(Brinkmann et al. 1997, Reeves \& Turner 2000).
Similar flat slopes for \X21 and \P2126 were reported  from previous 
ASCA and ROSAT
observations (Elvis et al. 1994{\bf b}, Cappi et al. 1997, Siebert et al. 1996)
and in a {\it Chandra} observation Fang et al. (2001) found an even 
flatter slope for \X21.

 In the frame of the two-emission components model 
 an  explanation  would be that the  emission is dominated by 
 the beamed, flat  blazar-like component.
The high X-ray luminosities of the 
two objects ($L_{\rm 2-10~keV}\simeq 2\times 10^{47}$ erg s$^{-1}$ and
$L_{\rm 2-10~keV}\simeq 7\times 10^{46}$ erg s$^{-1}$ for \P2126 and \X21, 
respectively) would be in accordance with this hypothesis.
 In the case of \P2126 we also find indication of beaming in the
extremely high radiative efficiency ($\eta \ma 8$) of accretion onto
a black hole exceeding the maximum theoretical value.

In order to get more insight into the properties of our sources
we calculated the broad band spectral indices 
between 5 GHz, 2500 \AA ~  and 2 keV, which are
good indicators of the SED's shape.
The values we obtained are $\alpha _{\rm ox}=1.11$, $\alpha _{\rm rx}=0.81$,
$\alpha _{\rm ro}=0.67$ for \P2126 , and $\alpha _{\rm ox}=0.95$,
$\alpha _{\rm rx}=0.81$, $\alpha _{\rm ro}=0.74$ for \X21. 
 Considering the uncertainty on the X-ray fluxes ($\la $ 30\%)
resulting from
the errors of the spectral fit parameters we estimate a typical
error for the $\alpha _{\rm ox}$ and $\alpha _{\rm rx}$ of the order of
${\mathbf \la 5}$\%. 
Padovani et al. (2002)  recently  studied a sample 
of FSRQ  characterized by a synchrotron peak at X-ray energies in the
same way as HBL BL Lacs. We checked if our sources could
belong to this class and thus explain their X-ray loudness. 
However, the above values of the broad band spectral indices place
our sources in the `normal' FSRQ region (see Fig. 1 of Padovani et al.
2002),
pointing at an inverse Compton origin for the X-ray emission from the
sources.
Indeed,  \X21 is known to have a blazar-like SED (Elvis et al. 2000)
with the synchrotron peak at $\sim 0.3$ mm and the inverse Compton
peak at $\sim 4$ MeV, providing strong support to its FSRQ classification.

On the other hand, \P2126 is known to be a GPS source (de Vries et al. 1997,
Stanghellini et al. 1998). These powerful radio sources are characterized
by extreme compactness ($\la 1$ kpc), low polarization and a convex
spectrum at radio frequencies with a turn-over between $\sim 500$ MHz
and $\sim 10$ GHz. This characteristic shape of the spectrum
is commonly interpreted in terms of synchrotron self-absorption 
(O' Dea 1998).
 Due to their double-lobed radio morphologies, GPS
sources are thought to be lying in the plane of the sky, so that beaming
shouldn't play any role. This makes the tentative classification 
of \P2126 as a FSRQ rather problematic  as already discussed by
Elvis et al. (1994b). However, new VLBA observations of
several GPS sources (Lister et al. 2002) revealed some atypical features 
for this class of objects,
like core-jet structures, super-luminal motion, variability and 
polarization  and this, together with the radiative efficiency 
argument (see above),
  suggest  relativistic
beaming at least for some of the  GPS sources. 
Moreover, these properties seem to agree better 
with a free-free absorption model from a surrounding ionized medium
than with the synchrotron absorption model (Lister et al. 2002).
Interestingly, this absorbing medium could also account for the extra
absorption we detect in \P2126 (see below).   

\subsubsection{Absorption properties}

Although excess absorption was previously reported (Cappi et al. 1997,
Siebert et al. 1996) for  \X21,  we  find no strong evidence for 
 extra absorption  for this source.
Depending on the assumed spectral form  the fits predict absorbing
column densities ranging from the galactic value to an additional
source intrinsic value of  $\Delta N_{\rm H,z}\simeq 1.1\times 10^{21}$ cm$^{-2}$
in case of a single power law fit over the whole energy range.
As the residuals of the fits indicate a more complex spectral
form than a simple power law the existence of any strong extra absorption must be
regarded as uncertain. 

For \P2126 we find additional absorption of  $\Delta N_{\rm H}\simeq 0.6\times 10^{21}$
cm$^{-2}$ in excess of the galactic value.
For this object extra absorption has been claimed before  from ASCA
and ROSAT observations (Elvis et al. 1994, Serlemitsos et al. 1994,
Cappi et al. 1997, Reeves \& Turner 2000). Their results are in
agreement with ours. 

The absence of extra absorption in \X21 argues against its ubiquity in
high redshift quasars (Yuan \& Brinkmann 1998).
 On the other hand, the presence of extra
absorption in \P2126 rises the question of the location of the
absorbing material. In fact, our fits are compatible with an absorber
at the redshift of the source, but other redshifts are equally
allowed by the data, so that this issue remains open.\\
A galactic origin of the extra absorption by means of molecular
clouds and dust has been excluded for this source by local CO
surveys and IRAS measurements (Cappi et al. 1997).
The absorption could then be due to intervening matter along the
line of sight. Damped Ly$\alpha $ systems with column densities
of the order of $10^{20}-10^{21}$ cm$^{-2}$ or intervening
galaxies could provide the necessary amount of absorption.
   
However, no damped Ly$\alpha $ systems have been detected in the
optical spectrum of \P2126, but only the  Ly$\alpha $ forest
(Giallongo et al. 1993) with much lower column densities
($N_{\rm H}\sim 10^{16}-10^{17}$ cm$^{-2}$), which cannot account
for the observed absorption. Moreover, some absorption variability
seems to be present in \P2126, not explicable by damped  Ly$\alpha $
systems, which are believed to be stable over long time scales.
 
It thus seems likely that the absorber is at the quasar's position.
Then a possible site for the absorber could be the torus invoked in the unified
model for AGNs or, allowing for variable absorption, dusty clouds.
A further alternative could be a cooling flow in a cluster of galaxies.
As \P2126 is classified as a GPS source, an interesting possibility is that
the proposed ionized medium (Lister et al. 2002) responsible for
the GPS phenomenon through free-free absorption, could be
the same absorbing medium we detect in the X-ray band.

 Our data do not allow to distinguish between a cold, neutral and a warm,
ionized absorber as  we cannot see in the spectrum the 
typical features of a warm absorber, i.e. the OVII and OVIII edges
at $\sim 0.7-1.0$ keV and an extra emission component below $\sim 0.7$
keV in the rest frame;  these energies are redshifted below
$\sim 0.2$ keV, outside the XMM band. 
  However, if the absorbing matter can be related to the 
Ly$\alpha$ clouds along the line of sight the temperature of the
gas as determined from high resolution optical observations 
(D'Odorico  et al. 1998) is around $T \sim 38000$K. 
The findings of Brinkmann et al. (1997) that the majority of  the 
high redshift objects in their sample are GPS sources
and that they show excess
absorption further supports the connection between the X-ray absorber and
the conceivable free-free absorber of Lister et al. (2002).

\begin{table*}
\small
\tabcolsep1ex
\caption{\label{summary} Summary of source properties. }
\begin{tabular}{lcccccccc}
\noalign{\smallskip} \hline \noalign{\smallskip}
\multicolumn{1}{c}{Source} & \multicolumn{1}{c}{type} &
\multicolumn{1}{c}{$z$} & \multicolumn{1}{c}{$m_{\rm V}$} &
\multicolumn{1}{c}{$\alpha _{\rm ox}$} &
\multicolumn{1}{c}{$\Gamma$} & \multicolumn{1}{c}{$L_{\rm 2-10~keV}$} &
\multicolumn{1}{c}{$N_{\rm H}$} & \multicolumn{1}{c}{$\chi^2_{\rm red}$/dof} \\
\multicolumn{1}{c}{  } & \multicolumn{1}{c}{ } & \multicolumn{1}{c}{ }
&\multicolumn{1}{c}{ } & \multicolumn{1}{c}{ (1) } &\multicolumn{1}{c}{  (2)  } &
\multicolumn{1}{c}{[$\times 10^{47}$ erg/s]} & \multicolumn{1}{c}{ (3)}
&\multicolumn{1}{c}{ } \\
\noalign{\smallskip} \hline \noalign{\smallskip}
 PKS 2149$-$306& RL  & 2.34 &
   18.5 & 0.95  & 1.53$\pm 0.02$ & 1.03 &  0.294 & 1.08/528 \\
 PKS 2126$-$158 & RL & 3.27 & 17.0 & 1.11 &
   1.51$\pm$0.02 & 1.97 & 1.1 & 0.98/492 \\
 Q 1442$+$2931& RQ  & 2.64 & 17.0 & 1.70
   & 1.92$\pm 0.07$ & 0.06 &  galactic & 0.87/105 \\
 Q 0000$-$263& RQ  & 4.10 &
   18.0 & 1.78 & 2.19$\pm 0.12$ & 0.05 &  galactic & 0.91/64 \\
\noalign{\smallskip}\hline
\end{tabular}
\medskip

 The errors given are at the 90\% level.\\
 (1): Estimated error of $\la 5$\%.\\
 (2): Power law slope  of fit to PN data with free absorption\\
 (3): Fitted  absorption in units of 10$^{21}$ cm$^{-2}$. \\
\end{table*}

\subsection{Radio-quiet objects}

A simple power law fit to the combined PN + MOS data of \Q0
gives a statistically acceptable representation of the spectrum and 
no absorption in excess of the galactic is found.
Unfortunately, the available number of photons is insufficient for 
a more complex modeling of the remaining residuals. 
Our findings are consistent with those from previous ROSAT observations
within the errors. We obtain from the combined PN+MOS fit a photon index of 
$\Gamma =2.10\pm 0.11$, 
whereas Bechtold et al. (1994) find  for the 0.1$-$ 2.4 keV band a spectral index 
$\alpha =1.30\pm 0.23$ and galactic absorption.
From this fit they deduced a broad spectral index $\alpha _{\rm ox}=1.85$
whereas we find a slightly different value   $\alpha _{\rm ox}=1.78$.
Kaspi et al. (2000) give a broad spectral index $\alpha _{\rm ox}=1.65$.
 The errors on the $\alpha _{\rm ox}$ have been estimated 
to be $\la 5$\% as in Sec.
\ref{411}.
The  discrepancies in these values  can be explained by the 
 different spectral slopes used in the
 calculations and the restricted ROSAT energy range compared to XMM.
 
Similar results hold for \1442, for which an acceptable fit is again
provided by a simple power law and galactic absorption. The power law
slope for the combined PN+MOS fit is $\Gamma =1.87\pm 0.05$. 
Due to the low number of counts no spectral analysis of ROSAT observations 
could be performed by Reimers et al. (1995) for a  
comparison with our results but the deduced fluxes are consistent
inside their errors. For this quasar we obtain a broad spectral
index $\alpha _{\rm ox}=1.70$.

The power law slopes we get for the two high redshift 
radio-quiet quasars are  in agreement 
with the typical values found at low redshifts (Yuan et al. 1998a),
suggesting the absence of spectral evolution for this class of AGN
and the presence of a constant spectral form over a wide band width.
Further, the slopes are also considerably steeper than those of the two radio-loud 
quasars of our sample, supporting the spectral dependence 
on the radio-loud/radio-quiet classification observed
at low redshifts (Brinkmann et al. 1997, Yuan et al. 1998a).

The $\alpha _{\rm ox}$ for these  two high redshift quasars are consistent
with values found in previous studies for $z>2$ and
they are larger than the average values found at $z<0.2$ (Yuan et al. 1998a).
This indicates that high redshift radio-quiet quasars are more X-ray quiet
than their low-redshift counterparts. However,  it is found
(Avni \& Tananbaum 1982, Avni \& Tananbaum 1986, Wilkes et al. 1994,
Yuan et al. 1998a)
that the X-ray loudness is independent of redshift and that instead a 
correlation is present between $\alpha _{\rm ox}$ and log $l_{\rm o}$, where $l_{\rm o}$
is the luminosity at 2500 \AA  . Thus the larger $\alpha _{\rm ox}$ 
 would arise from a higher $l_{\rm o}$ at high redshift for these 
optically selected quasars.
From the broad spectral indices it is also inferred
that radio-loud quasars are X-ray louder than radio-quiet quasars  
at high redshifts (Brinkmann et al. 1997)  as well as
at low redshifts (Zamorani et al. 1981). This is supported
by our data, suggesting that an additional component contributes to
the X-ray emission in radio-loud quasars and that radio-quiet and radio-loud 
quasars have distinct physical emission mechanisms.   
However the `X-ray quietness' of radio-quiet quasars makes a detailed
spectral analysis rather difficult, and the small number of well
studied radio-quiet high-z quasars  make definite conclusions rather 
uncertain. 

No excess absorption has been found in the radio-quiet objects
confirming that this property is  common only in the
high redshift, radio-loud quasars, even if not ubiquitous.
A larger sample of high redshift radio-quiet quasars is needed
 to study this issue properly.
As already mentioned for \P2126, damped Ly$\alpha $ systems
have been discussed as possible X-rays absorbers for radio-loud
high redshift quasars.
Since its discovery \Q0 is known to have a damped Ly$\alpha $ system
lying along its line of sight at $z=3.39$ with   $N_{\rm HI}\sim 2.6\times 10^{21}$
cm$^{-2}$ (Levshakov et al. 2000). 
 We added the above fixed  amount of absorbing material at the 
 Ly$\alpha $ system's redshift  to the power law fit of \Q0 and could not
find any statistically significant differences in the fit parameters  
from a fit with galactic absorption only. The main differences in the two models
occur at lowest energies ($\leq 0.3 $ keV)  where the PN is not sensitive 
enough. As the count rates for \Q0 are quite low, especially for the RGS,
 this means 
that the data are insufficient to determine
the amount of absorbing material at high redshift for this source.
 
\section{Summary} 

A brief summary of source properties and results of the spectral analyses are
given in Table \ref{summary}.
Please note that we give, for a comparison,  the power
law slope for the   PN fit with free absorption; the resulting 
$\chi^2_{\rm red}$, 
given in the last column, indicates the quality of the fit.
A simple power law fit for PKS~2149-306 with $\Gamma \sim 1.53$ and
$N_{\rm H}=2.94\times 10^{20}$ cm$^{-2}$, slightly in excess of the
galactic value, provides  an acceptable description of the data 
in the $0.2-10$ keV energy band.
Allowing for extra cold or warm absorption at the redshift of the 
source results
in  equally acceptable fits with an identical slope and
$N_{\rm H,z}\sim 10^{21}$ cm$^{-2}$. However the $N_{\rm H,z}$ is
ill-determined and other redshifts of the absorber are compatible
with the data. 
With the high  signal to noise ratio from the EPIC data there is evidence 
for substantial deviations
from a simple power law, such as a slightly curved shape of the
residuals at soft energies, which can be modeled as a black body component
with $kT\sim 0.25$ keV, different power law slopes when different
energy ranges are used for the fits, and the flatter slopes found from
the MOS fit. We conclude that there is no strong evidence for extra absorption
for this quasar. 
The addition of a narrow line at $\sim 5$ keV as claimed by Yaqoob et al. 
(1995) with a rest frame equivalent width of $\sim 12$ eV is not significant.
Only a further improved signal to noise ratio would allow
to separate different components in the spectrum.

For  \P2126 extra absorption was found with 
$N_{\rm H}=1.1\times 10^{21}$ cm$^{-2}$ and a power law slope of
$\Gamma \sim 1.51$. In fits with  extra absorption in the source frame, 
either neutral or through a warm absorber,  values of 
$N_{\rm H,z}\sim  10^{22}$ cm$^{-2}$ and 
$\Gamma \sim 1.47$ were found. However the redshift of the absorbing
medium is poorly constrained. Further, deviations from the
simple power law which cannot be modeled properly due to
the still insufficient number of detected photons, seem to be present.
A higher signal to noise ratio is needed,  both to study
the presence of the various spectral components and to determine
the redshift of the absorber.

Both radio-loud quasars have high $2-10$ keV luminosities of
the order of $\sim 10^{47}$ erg s$^{-1}$ and they are X-ray loud
with $\alpha _{\rm ox}\sim 1$.
The X-ray emission of \X21 is dominated by 
a beamed, flat blazar-like component produced by inverse
Compton scattering; the shape of the SED similar to that of
a blazar confirms this view.

\P2126 is a GPS source and the absorber we detect in X-rays 
might be  the same as that assumed
in the free-free absorption model for GPS sources (O' Dea 1998,
Lister et al. 2002).
 From  its spectral properties, \P2126 seems
to have blazar-like characteristics  similar to those of
 \X21, apparently in contrast with
being a GPS source, with a jet oriented at a large angle
to the line of sight. However, some GPS sources exist which
show signs of relativistic beaming as  typically found in blazars
(Lister et al. 2002).

For both radio-quiet objects a simple power law with 
$\Gamma \sim 2$ and 
galactic absorption gives a good description of the data.
The power  law slopes we found are consistent with typical values
at low redshifts.
No iron lines have been detected. 
The $2-10$ keV luminosities are of the order of
$\sim 10^{45}$ erg s$^{-1}$, much lower than for their radio-loud
counterparts. The  $\alpha _{\rm ox}$ of about $\sim 1.7$ are
considerably larger than for the two radio-loud quasars.
Being much X-ray weaker 
than radio-loud quasars, the low signal to noise ratios inhibit the 
detection of   possible deviations from a simple power law slope.

The two radio-loud objects are found in the upper - left
region of the $\alpha_{\rm ro} - \alpha_{\rm ox}$ diagram, slightly
offset from the  average of the low-z radio-loud quasars,
similar to the $z>4$ quasars of Fabian et al. (1999).
They are thus X-ray brighter than their low-z counterparts and
follow the $ \alpha_{\rm ox}$ - $l_o$ relation of radio-loud quasars
(Brinkmann et al. 1997).
The two radio-quiet objects are significantly more X-ray quiet
than those at low redshifts and their  $ \alpha_{\rm ox}$ and their
spectral power law slopes are nicely in line with the
redshift dependence of these objects (Bechtold et al. 2002).

Overall, our sources follow the general trends observed for other
high-z quasars.
However,  in contrast to previous work, the spectral parameters could be
determined with much higher accuracy. 
 The great advantage  of the XMM-Newton instruments for studies of
high redshift quasars is the large sensitivity of the instrument.
This allows not only the detection but even a spectral study of these
 objects  and,
in  the case of more luminous sources, to  study small
spectral  deviations from a simple power law which are indicators
of the physical conditions governing the
emission of radiation over the wide energy band accessible
in high-z objects.

\vskip 0.4cm
\begin{acknowledgements}
This research has made use of the NASA/IPAC Extragalactic Data Base
(NED) which is operated by the Jet Propulsion Laboratory,
California Institute of Technology, under contract with the National
 Aeronautics and Space Administration.
This work is based on observations with XMM-Newton, an ESA science mission 
with instruments and contributions directly funded by ESA Member States
 and the USA (NASA).
\end{acknowledgements}


\begin{thebibliography}{}
 \bibitem{} Avni, Y., \& Tananbaum, H. 1982, ApJ, 262, L17
 \bibitem{} Avni, Y., \& Tananbaum, H. 1986, ApJ, 305, 83
\bibitem{} Ballantyne, D.R., Iwasawa, K., \& Fabian, A.C. 2001, MNRAS, 323, 506 
\bibitem{} Bechtold, J., Elvis, M., Fiore, F., Kuhn, O., Cutri, Roc M., Czerny, B., Janiuk, A., et al. 1994a, AJ, 108, 374
\bibitem{} Bechtold, J., Siemiginowska, A., Shields, J., et al. 2002, astro-ph/0204462
\bibitem{} Brinkmann, W., Siebert, J., Reich, W., Fuerst, E., Reich, P., et al 1995,
   A\&AS,  109, 147
\bibitem{} Brinkmann, W., Yuan, W., \& Siebert, J. 1997, A\&A, 319, 413
\bibitem{} Boyle, B.J., Griffiths, R.E., Shanks, T., Stewart, G.C., \& Georgantopoulos, I.  1993, MNRAS, 260, 49
\bibitem{} Cappi, M., Matsuoka, M., Comastri, A., Brinkmann, W., Elvis, M., et al. 1997, ApJ, 478, 492
\bibitem{} de Vries, W.H., O'Dea, C. P., Baum, S.A., \& Barthel, P.D. 1997, AAS, 191.2204D
\bibitem{} D'Odorico, V., Christiani, S., D'Odorico, S., Fontana, A., \& Giallongo, E.
 1998, A\&AS 127, 217
\bibitem{} Ehle, M., Breitfellner, M., Dahlem, M., Guainazzi, M., Rodriguez, P., et al. 2001,
 XMM-Newton Users' Handbook, \hfill \break
          $http://xmm.vilspa.esa.es/xmm\verb"_"user\verb"_"support/$
         \hfill \break $external/documentation/uhb\verb"_"frame.shtml$
\bibitem{} Elvis, M., Wilkes, B.J., McDowell, J.C., Jonathan, C., Green, R.F.,
et al. 1994a, ApJS, 95, 1
\bibitem{} Elvis, M., Fiore, F., Wilkes, B.J., McDowell, J., \& Bechtold, J. 1994b, ApJ, 422, 60
\bibitem{} Elvis, M., Fiore, F., Siemiginowska, A., Bechtold, J., Mathur, S., et al. 2000, ApJ, 543, 545
\bibitem{} Fabian, A.C. 1979, Proc. R. Soc. London, Ser.A, 366, 449
\bibitem{} Fabian, A.C., Celotti, A., \& Pooley, G. 1999, MNRAS, 308, L6
\bibitem{} Fang, T., \&  Canizares, C. 2000, ApJ, 532, 539
\bibitem{} Fang, T., Marshall, H.L., Bryan, G.L., \& Canizares, C. 2001, ApJ, 555, 536
\bibitem{} Ghizzardi, S., \& Molendi, S. 2001, Proc. of the conference
   'New Visions of the X-ray Universe', ESTEC Nov. 2001
\bibitem{} Giallongo, E., Cristiani, S., Fontana, A., \& Trevese, D. 1993, ApJ, 416, 137
\bibitem{} Kaspi, S., Brandt, W.N., \& Schneider, D.P. 2000, AJ, 119, 2031 
\bibitem{} Lister, M.L., Kellermann, K.I., \& Pauliny-Toth, I.I.K. 2002, astro-ph/0207175
\bibitem{} Levshakov, S.A., Molaro, P., Centurion, M., D'Odorico, S., Bonifacio, P., et al. 2000, A\&A, 361, 803
\bibitem{} Mushotzky, R.F., Done, C., \& Pounds, K.A. 1993, ARA\&A, 31, 717
\bibitem{} O'Dea, C.P. 1998, PASP, 110, 493
\bibitem{} Padovani, P., Costamante, L., Ghisellini, G., Giommi, P., \& Perlman, E.  2002, astro-ph/0208501
\bibitem{} Reimers, D., Bade, N., Schartel, N., Hagen, H.-J., Heber, U., et al. 1995, A\&A, 296, L49
\bibitem{} Reeves, J.N., \&  Turner,  M.J.L. 2000, MNRAS, 316, 234
\bibitem{} Sanduleak, N., \& Pesch, P. 1989, PASP, 101, 1081
\bibitem{} Schartel, N., Walter, R., Fink, H.H., \& Tr\"umper, J. 1996,
  A\&A, 307, 33
\bibitem{} Schartel, N., Komossa, S., Brinkmann, W., Fink, H.H., Tr\"umper, J., et al. 1997, A\&A, 320, 421
\bibitem{} Serlemitsos, P., Yaqoob, T., Ricker, G., Woo, J., Hideyo, K., et al. 1994, PASJ, 46, L43
\bibitem{} Siebert, J.,  Matsuoka, M., Brinkmann, W., Cappi, M., Mihara, T., et al. 1996, A\&A, 307, 8
\bibitem{} Sikora, M., Madejski, G., Moderski, R., \& Poutanen, J.  1997, ApJ, 484, 108
\bibitem{} Stanghellini, C., O'Dea, C.P., Dallacasa, D., Baum, S.A., Fanti, R., et al. 1998, A\&AS, 131, 303
\bibitem{} Wilkes, B.J., \& Elvis, M. 1987, ApJ, 323, 243
\bibitem{} Wilkes, B.J., Tananbaum, H., Worrall, D.M., Avni, Y., Oey, M.S., et al. 1994, ApJS, 92, 53
\bibitem{} Worrall, D.M., \& Wilkes, B.J. 1990, ApJ, 360, 396
\bibitem{} Yaqoob, T., George, I.M., Nandra, K., Turner, T.J., Zobair, S., et al. 1999, ApJ, 525, L9
\bibitem{} Yuan, W., Brinkmann, W., Siebert, J., \& Voges, W. 1998a, 
 A\&A,  330, 108 
\bibitem{} Yuan, W., \& Brinkmann, W. 1998, in "Highlights in X-ray Astronomy",
 eds B. Aschenbach \& M.J. Freyberg, MPE Report No. 272, 240
\bibitem{} Yuan, W., Siebert, J., \& Brinkmann, W. 1998b, A\&A, 334, 498
\bibitem{} Yuan, W., Matsuoka, M., Wang, T., Ueno, S., Kubo, H., et al. 2000, ApJ, 545, 625
\bibitem{} Zamorani, G., Henry, J.P., Maccacaro, T.,  Tananbaum, H., Soltan, A., et al.  1981, ApJ, 245, 357
\end{thebibliography}
\end{document}